\documentclass[a4paper, 11pt]{article}
\usepackage{jcappub} 
\RequirePackage[top=28mm,bottom=28mm,left=22mm,right=22mm]{geometry}

\usepackage{enumitem}
\usepackage{fancyhdr}
\usepackage{mathtools}
\usepackage{amssymb}
\usepackage{amsmath}
\usepackage{hyperref}
\usepackage{amsfonts}
\usepackage{amssymb}
\usepackage{caption}
\usepackage{float}
\usepackage{bm}
\usepackage{graphicx}
\usepackage{color}
\usepackage{verbatim}
\usepackage{blindtext, multicol}
\usepackage{multirow}
\usepackage{epigraph}
\usepackage[dvipsnames]{xcolor}
\usepackage{physics}

\usepackage[normalem]{ulem}

\usepackage{tensor}
\usepackage{physics}
\usepackage{romannum}
\usepackage{subfigure}

\usepackage[sorting=none,style=numeric-comp]{biblatex}
\usepackage{academicons}
\usepackage{fontawesome5}
\definecolor{orcidlogocol}{rgb}{0.65, 0.807, 0.223}

\newcommand{\orcid}[1]{$\,$\href{https://orcid.org/#1}{\textcolor{orcidlogocol}{\faOrcid}}}

\numberwithin{equation}{section}

\def\beq{\begin{equation}}
\def\eeq{\end{equation}}
\def\ber{\begin{eqnarray}}
\def\eer{\end{eqnarray}}
\def\benu{\begin{enumerate}}
\def\eenu{\end{enumerate}}

\def\l{\left}
\def\r{\right}
\def\d{{\rm d}}
\newcommand{\sq}{\lower.25ex\hbox{\large$\Box$}}

\def\f{\frac}
\def\mpl{m_{p}}

\def\Og{\Omega_{_{\rm GW}}}

\def \t {\times}

\def \red {\color{red}}
\def \blue {\color{blue}}
\def \Blue {\color{Blue}}
\def \fgreen {\color{ForestGreen}}

\def \violet {\color{violet}}

\def \Brown {\color{Brown}}

\def \borange {\color{BurntOrange}}

\def \half {\frac{1}{2}}

\hypersetup{
   colorlinks = true,
   linkcolor = blue,
   citecolor = Brown,
   urlcolor = Blue
}
\usepackage{enumitem}

\usepackage{cancel}

\usepackage{framed}

\colorlet{shadecolor}{CadetBlue!5}

\title{{\Brown Morphology of Inflationary Gravitational Wave Spectra imprinted by a Sequence of Post-Inflationary Epochs \textit{via} \texttt{GWInSpect}}}
\author[a,b]{{\Blue Swagat~S.~Mishra}\,\orcid{0000-0003-4057-145X},}
\author[c]{{\Blue Athul~K.~Soman}\,\orcid{0009-0002-0947-6892},}
\affiliation[a]{School of Physics and Astronomy,  University of Nottingham, Nottingham, NG7 2RD, UK.}
\affiliation[b]{Cosmology, Gravity, and Astroparticle Physics Group, Center for Theoretical Physics of the Universe (CTPU-CGA), Institute for Basic Science (IBS), Daejeon, 34126, Korea.}
\affiliation[c]{International School for Advanced Studies (SISSA), via Bonomea 265, 34136 Trieste, Italy.}
\emailAdd{swagat.mishra@nottingham.ac.uk}
\emailAdd{akuruvai@sissa.it}

\abstract{Expansion history of the Universe, prior to the onset of Big Bang Nucleosynthesis (BBN), remains largely unknown. The high-energy post-inflationary era is, in general, expected to be complex, potentially consisting of multiple distinct epochs, each characterized by a distinct equation of state (EoS). One of the robust predictions of the inflationary paradigm is the generation of tensor perturbations through quantum fluctuations, which later manifest as a stochastic background of primordial gravitational waves (GWs). The large-scale amplitude and small-scale spectral tilt ($n_{\rm GW}$)  of inflationary GWs encode information about the energy scale of inflation, and provide an important observational probe of the post-inflationary (pre-BBN) dynamics, respectively. In particular, a softer post-inflationary EoS ($w < 1/3$) leads to  red-tilted GW spectrum ($n_{\rm GW} < 0$), whereas  stiffer EoS ($w > 1/3$) produces  blue tilt ($n_{\rm GW} > 0$). 

In the previous work~\cite{Soman:2024zor}, we developed an analytical framework for computing the spectral energy density of these first-order GWs for scenarios involving multiple sharp (instantaneous) transitions in the post-inflationary EoS, $w_1 \to w_2 \to \cdots \to w_n \to w_{n+1} = 1/3$. We primarily focused on determining the parameter space which leads to GW signals that may be detectable by future GW observatories. 

In this companion paper, we extend the framework of Ref.~\cite{Soman:2024zor} to systematically explore the rich landscape of possible inflationary GW spectra imprinted by multiple  ($n \gg 1$) post-inflationary epochs prior to the hot Big Bang. Remaining agnostic to specific models, we demonstrate that a {\em diverse morphological zoo of spectral shapes}\,--\,ranging from convex and concave monotonic profiles to highly non-monotonic features\,--\,can emerge depending on the sequence and duration of epochs. We introduce \texttt{GWInSpect}\,\href{https://github.com/athul104/GWInSpect.git}{\faGithub}~--~a Python-based, user-friendly package that efficiently generates inflationary GW spectra for arbitrary post-inflationary histories, enabling a systematic study of the post-inflationary expansion history. \\
}
\arxivnumber{arXiv:2510.25672}

\begin{document}

\maketitle
\flushbottom

\section{Introduction}
\label{sec:Intro}
Cosmic inflation~\cite{Starobinsky:1980te,Guth:1980zm,Linde:1981mu,Albrecht:1982wi,Linde:1983gd,Linde:1990flp,kinney2009tasi,Martin:2013tda,Baumann_TASI,Baumann:2018muz,Kodama1984_CPT,Riotto2002_CPT,Ellis:2023wic,Mishra:2024axb} is the leading paradigm for providing initial conditions for the hot Big Bang phase~\cite{Linde:1990flp,Kolb:1990vq,Dodelson:2003ft,Rubakov:2017xzr}. In its simplest realization, inflation is driven by a scalar field $\phi$, the \textit{inflaton},  which is minimally coupled to gravity, and slowly rolls down its potential $V(\phi)$~\cite{Linde:1990flp,Martin:2013tda,Baumann_TASI,Mishra:2024axb}. Quantum fluctuations of the inflaton field give rise to a nearly scale-invariant spectrum of scalar perturbations~\cite{Mukhanov:1981xt,Hawking:1982cz,Starobinsky:1982ee,Guth:1982ec}, whose imprints on the Cosmic Microwave Background (CMB)~\cite{Planck:2018vyg,Planck_overview,Planck_inflation,2021PhRvL.127o1301A} and the large-scale structure~\cite{Dodelson:2003ft,Mukhanov:2005sc,Baumann:2022mni} strongly support the single-field slow-roll inflationary framework.

Alongside scalar perturbations, inflation also generates tensor quantum fluctuations, which constitute a stochastic background of primordial gravitational waves (GWs)~\cite{Grishchuk:1974ny,Starobinsky:1979ty,Sahni:1990tx,Allen:1987bk}. Their spectral energy density ($\Omega_{\rm GW}$) on large cosmological scales contains information about the energy scale of inflation~\cite{Starobinsky:1979ty,Sahni:1990tx,Giovannini:1999hx,Mishra:2021wkm,Haque:2021dha,Soman:2024zor}, while  its spectral tilt ($n_{\rm GW}$), particularly   on small scales, is sensitive to the post-inflationary equation of state (EoS), $w$, of the early Universe~\cite{Sahni:1990tx,Soman:2024zor}. To be precise, a softer  EoS, with $w<1/3$, results in a red tilt, while a stiffer  EoS, with $w > 1/3$, leads to blue-tilted GWs, which are of great phenomenological interest from an observational prospect. Consequently, the inflationary GW spectrum offers a unique window into the expansion history of the Universe between the end of inflation and the onset of Big Bang Nucleosynthesis (BBN)\,--\,which remains observationally unconstrained at present~\cite{Giovannini:2019oii,Caprini:2018mtu,Guzzetti:2016mkm,Wang:2016tbj,Soman:2024zor,Mishra:2024axb}.

A broad network of GW detectors, operating across frequencies from $10^{-18}$~Hz to $10^3$~Hz, is now capable of probing this primordial signal. These include CMB B-mode experiments such as \texttt{BICEP/Keck}~\cite{BICEP:2018czh}, \texttt{LiteBIRD}~\cite{Hazumi:2019lys}, and the \texttt{Simons Observatory}~\cite{POLARBEAR:2015ixw}; pulsar timing arrays (PTAs) such as \texttt{NANOGrav}~\cite{NANOGrav:2023gor,NANOGrav:2023tcn}, \texttt{EPTA/InPTA}~\cite{EPTA:2023fyk,EPTA:2023sfo}, and \texttt{IPTA}~\cite{InternationalPulsarTimingArray:2023mzf}; and laser interferometers such as \texttt{LIGO/Virgo/KAGRA}~\cite{LIGOScientific:2014pky,VIRGO:2014yos,KAGRA:2018plz} and future missions including \texttt{LISA}~\cite{LISA:2017pwj}, \texttt{DECIGO}~\cite{Kawamura:2019jqt}, \texttt{BBO}~\cite{Harry:2006fi}, \texttt{Cosmic Explorer}~\cite{Reitze:2019iox}, and the \texttt{Einstein Telescope}~\cite{Coccia:2023wag}. These facilities together span a significant frequency range of inflationary tensor modes.

The post-inflationary Universe is expected to have undergone a complex sequence of phases and non-equilibrium processes before reaching the radiation-dominated phase~\cite{Kofman:1997yn,Lozanov:2019jxc,Amin:2014eta,Allahverdi:2020bys} prior to the commencement of BBN. The dynamics of reheating and subsequent transitions can, therefore, involve several distinct epochs~\cite{Giovannini:2022vha,Haque:2021dha,Gouttenoire:2021jhk,Gouttenoire:2019kij,Gouttenoire:2019rtn,Maiti:2025ijr,Maiti:2025cbi}, each characterized by a different effective EoS, $w_i$. Most works in the literature have typically assumed a single average EoS during reheating~\cite{Martin:2014nya,Figueroa:2019paj,Vagnozzi:2020gtf,Benetti:2021uea,Vagnozzi:2023lwo}.
In our previous work~\cite{Soman:2024zor}, we developed an analytical framework based on Israel junction-matching conditions~\cite{Deruelle:1995kd,Sahni:1990tx,Mishra:2023lhe} to compute the  spectrum of the first-order inflationary GWs for multiple ($n$-number of) sharp (instantaneous) transitions in the post-inflationary EoS, $w_1 \to w_2 \to \cdots \to w_n \to 1/3$. We primarily focused on identifying the parameter space of post-inflationary EoS  and transition energy scales, $\lbrace w_i,\,E_i \rbrace$, that leads to the spectra of inflationary GWs which are consistent with the existing constraints from \texttt{aLIGO} and BBN, while being potentially detectable by future observatories. In fact, Ref.~\cite{Soman:2024zor} dealt with up to $4$ different post-inflationary epochs before the hot Big Bang.
\begin{shaded}
In this companion paper, we utilise the framework developed in Ref.~\cite{Soman:2024zor} to systematically characterise the \textit{morphological zoo of  inflationary GW spectra} emerging from a large number of sequential post-inflationary epochs ($n \approx 10$). We  remain agnostic about the specific model details, and allow the EoS of any epoch to lie in the range $w_i \in \l( -1/3, \,1 \r)$. We demonstrate that a remarkably diverse range of spectral shapes\,--\,convex, concave, monotonic, and non-monotonic\,--\,can arise purely from the sequence and duration of these epochs. We further develop a simple and user-friendly \texttt{Python}-based {\em  numerical package}, called \texttt{\Brown \bf GWInSpect} (\texttt{GitHub} link \href{https://github.com/athul104/GWInSpect.git}{\faGithub}) that facilitates the interested users to generate such spectra for arbitrary post-inflationary histories within this general framework.  In fact, one of the key objectives of this work is to highlight the efficiency of \texttt{GWInSpect} package. With a simple interface and fast performance, it enables the exploration of complex sequences of epochs with minimal effort, making it ideal for large parameter scans or integration into broader cosmological analyses. We provide an instructive tutorial with an \texttt{ipynb} notebook in the same \texttt{GitHub} page, which can be directly accessed from the link\,:~\href{https://github.com/athul104/GWInSpect/blob/main/examples/tutorial_gwinspect.ipynb}{{\bf \blue tutorial notebook}}.  By providing this package publicly, we positively aim to encourage its widespread use for systematically investigating the imprints of diverse post-inflationary scenarios on the gravitational wave background. 
\end{shaded}
To be precise, the present paper differs from our previous work in Ref.~\cite{Soman:2024zor} in the following distinct ways\,:
\begin{itemize}
    \item The previous work in Ref.~\cite{Soman:2024zor} presented a comprehensive analytical and numerical framework for computing the spectral energy density of inflationary GWs, corresponding to multiple post-inflationary epochs in the early Universe. The primary goal was to determine the viable parameter space of the post-inflationary Universe\,--\,in particular, the equation-of-state  parameters and the energy scales of successive transition epochs\,--\,that could yield an observable GW signal in upcoming and future detectors, while remaining consistent with existing constraints from BBN and \texttt{aLIGO}. 
    
    In the present work, we instead utilise the analytical formalism developed in Ref.~\cite{Soman:2024zor}  to systematically categorise the wide variety of inflationary GW spectra that can emerge from multiple EoS transitions. While all our results respect the BBN constraints, we do not focus here on detectability forecasts or comparison with experimental sensitivities.  In contrast, the present paper adopts a narrower focus, emphasizing the diversity of possible spectral shapes arising from multiple post-inflationary epochs. Since our analytical approach relies on the instantaneous transition approximation between successive epochs, each post-inflationary phase is required to last at least one $e$-fold for a reliable estimate of the GW spectrum.

    \item Ref.~\cite{Soman:2024zor} applied the formalism to  model-agnostic post-inflationary histories, as well as to a specific string theory inspired scenario. The current work, on the other hand, is fully model-independent and aims to illustrate the generic features of the inflationary GW spectrum without reference to any particular particle-physics model.

    \item Building upon the numerical infrastructure developed previously, we introduce a dedicated publicly available numerical package, \texttt{GWInSpect}, which efficiently computes the inflationary GW spectrum for arbitrary sequences of post-inflationary epochs characterized by multiple EoS parameters $\{w_i\}$. The code allows the user to generate spectra across a wide frequency range within a short runtime. 
    
\end{itemize}

The paper is organized as follows. In Sec.~\ref{sec:Omega_GW}, we summarize the relevant analytical expressions for the tensor mode functions in the post-inflationary epochs, and the corresponding GW spectral energy density, using the junction matching formalism discussed in Ref.~\cite{Soman:2024zor}. In Sec.~\ref{sec:Inf_GW_Zoo}, we present the wide variety of possible spectral shapes of inflationary GWs corresponding to different post-inflationary expansion histories. Sec.~\ref{sec:Num_GWInSpect} describes our lightweight \texttt{Python}-based numerical package, \texttt{GWInSpect}, for computing these spectra efficiently. Finally, Sec.~\ref{sec:Discussions} summarizes our main conclusions and provides further discussion.

\begin{shaded}
\vspace{-0.25in}
\begin{center}
{\bf \Brown Units, Notation \& Convention}
\end{center}
\begin{itemize}
\item We mostly work in  natural units, where ~$\hbar,c =1$.
\item The reduced Planck mass is defined to be ~$m_p = \f{1}{\sqrt{8\pi G}} = 2.44 \times 10^{18}$\,GeV.
\item The background universe is considered to be described by the flat Friedmann-Lema\^{i}tre-Robertson-Walker (FLRW) line element, with scale factor $a(t)$  and Hubble parameter $H(t)$. 
\item Conformal time ($\tau$) is related to the cosmic time ($t$) by ${\rm d} \tau =\f{{\rm d}t}{a(t)}$.
\item Fourier mode functions $h_k(\tau)$ of a field $h(\tau,\vec{x})$ are defined as
$$ h(\tau,\,\vec{x}) = \int \f{\d ^3 \vec{k}}{\l(2\,\pi\r)^3} \, h_k(\tau)\,e^{i\,\vec{k}.\vec{x}}\, .$$
\item Derivative of the mode functions \textit{w.r.t} cosmic time $t$ is denoted as an {\em overdot}, $\frac{\d h_k}{\d t} \equiv \dot{h}_k$ and \textit{w.r.t} conformal time  $\tau$ is denoted as an {\em overprime}, $\frac{\d h_k}{\d \tau} \equiv h^\prime_k$.
\item The dimensionless Hubble constant, denoted as $h$, is defined by $h = H_0/ \l[ 100\,\l({\rm km/sec/Mpc}\r)\r]$ [\,not to be confused with the tensor field $h(\tau, \vec{x})$, or the tensor mode functions $h_k(\tau)$\,]. 
\end{itemize}
To remain consistent, we follow the same notation and convention as used in Ref.~\cite{Soman:2024zor}.
\end{shaded}

\section{Inflationary gravitational waves}
\label{sec:Omega_GW}

The accelerated expansion of space during inflation leads to the amplification of vacuum tensor fluctuations~\cite{Grishchuk:1974ny}, which are stretched to super-Hubble scales~\cite{Starobinsky:1979ty}, where they remain effectively frozen until their subsequent Hubble re-entry in the post-inflationary (decelerating) Universe. Once these tensor modes re-enter the Hubble radius, they begin to oscillate and propagate as primordial GWs. Modes with higher frequencies (shorter wavelengths) re-enter the Hubble radius at earlier times after inflation, whereas modes with lower frequencies (longer wavelengths) re-enter at later epochs. Having been generated during inflation, the tensor modes propagate largely unimpeded to the present epoch, thereby retaining key imprints of the intermediate expansion history of the Universe. In particular, the high-frequency regime of the spectrum, corresponding to present-day frequencies $f \gg 10^{-10}\,{\rm Hz}$, encodes valuable information about the unknown pre-BBN evolution of the Universe (see Sec.~\ref{sec:Omega_GWs_postinf}), as discussed in Refs.~\cite{Sahni:1990tx,Figueroa:2019paj,Soman:2024zor}. In what follows, we summarise the essential analytical framework required for computing the spectral energy density of inflationary GWs, restricting ourselves to those expressions most relevant for the present study. A detailed  treatment of the same can be found in our previous work~\cite{Soman:2024zor}.

\subsection{Evolution of tensor mode functions through  multiple post-inflationary epochs}
\label{subsec:tensor_postinf}
Evolution of the Fourier mode functions of primordial tensor fluctuations generated by inflation, at linear order in perturbation theory,  is governed by the following second-order differential equation 
\begin{equation}\label{eq:EOM_hk}
    h_k ^{'' \lambda}  + 2 \, \l( \frac{a'}{a} \r) \, h_k ^{' \lambda}  + k^2  \, h_k ^{\lambda}= 0 \, ,
\end{equation}
where $h_k ^{\lambda} $ is the tensor amplitude, with $\lambda = \lbrace +,\,\times \rbrace$ denoting the two polarisation of GWs and $k$ is the comoving wavenumber of the tensor modes. General  solution to Eq.~\eqref{eq:EOM_hk} can be written as a linear combination of two Bessel functions of the first kind~\cite{NIST:DLMF},  namely, $\lbrace J_{ \l(\alpha - \half \r)} (x), \, J_{- \l(\alpha - \half \r)}(x)\rbrace$, as follows\,--
\begin{equation}
    h_k ^{\lambda} \l(y_{_k}\r) = \frac{1}{(\alpha \, y_{_k})^{\alpha - \half}} \l[A_k \, J_{ \l(\alpha - \half \r)} (\alpha \, y_{_k}) +  B_k \, J_{- \l(\alpha - \half \r)} (\alpha \, y_{_k}) \r] \label{eq:General_exp_for_hk} \; ,
\end{equation}
where $\alpha$ is related to the EoS ($w$), while $y_{_k}(\tau)$ is the physical frequency ($k/a(\tau)$) of the tensor modes\,---\,scaled by the Hubble parameter ($H(\tau)$),  of the post-inflationary Universe, \textit{i.e.},
\beq
\alpha = \f{2}{1+3\,w} \, , \quad y_{_k} (\tau) = \f{k}{a(\tau) H(\tau)} \, .
\label{eq:alpha_w_relation}
\eeq
 The Bogoliubov coefficients, $\lbrace A_k,\,B_k\rbrace$, are determined by imposing the appropriate initial or boundary conditions on the tensor mode functions. For the first post-inflationary epoch, characterised by a constant EoS ($w_1$), these coefficients are obtained by matching to the inflationary tensor modes at the end of inflation. Specifically, the super-Hubble inflationary tensor perturbations serve as the initial conditions for the subsequent evolution in this epoch. Consequently, for any mode with comoving wavenumber $k$ that was super-Hubble at the onset of the first epoch, the initial tensor mode function is taken to be equal to the frozen amplitude of the corresponding inflationary mode, while its time derivative vanishes, namely,
\begin{equation}\label{eq:hk_initial_condition}
        h_k ^{\lambda} (\tau_k) =   h_{k, \, {\rm inf}} ^{\lambda} \quad ; \quad h_k ^{' \lambda} (\tau_k) = 0 \, ,
\end{equation}
where  $\tau_k$ marks the post-inflationary Hubble-entry time of a tensor mode of interest with (comoving) wavenumber $k$. Eq.~\eqref{eq:hk_initial_condition} yields the following expression for the Bogoliubov coefficients for the first epoch (see Ref~\cite{Soman:2024zor})
\begin{equation}\label{eq:coeff_Ak_Bk_first_epoch}
    A_{k, \, 1} = 2^{\l(\alpha_1 - \frac{1}{2}\r)} \,   \Gamma \l(\alpha_1 + \f{1}{2}\r) \,  h_{k, \, {\rm inf}} ^{\lambda}, \quad B_{k, \, 1} = 0 \, ,
\end{equation}
where $\Gamma(x)$ is the Gamma function. 

\begin{figure}[htb] 
    \centering
    \includegraphics[width = 1.0 \textwidth]{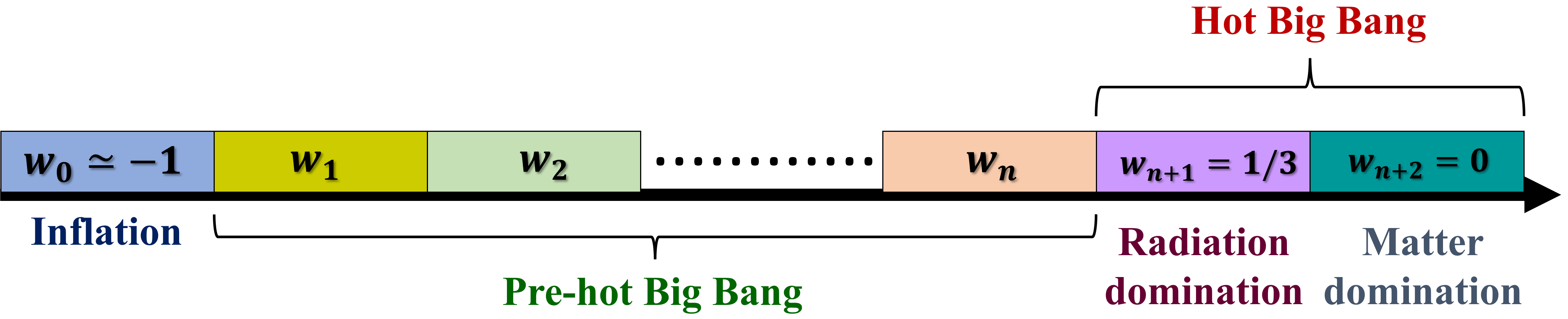}
    \caption{A schematic depiction of the timeline of the universe featuring multiple  post-inflationary epochs, each described by a nearly constant equation of state parameter. Note that the horizontal length is not a representation of the actual duration.}
    \label{fig:Expansion_EoS}
\end{figure}

As emphasized earlier, the high-energy post-inflationary dynamics of the Universe is expected to involve a variety of complex physical processes, potentially giving rise to multiple distinct expansion phases before the onset of BBN. It is therefore instructive to model the pre-BBN expansion history as a sequence of successive cosmic epochs, each characterised by a (nearly) constant EoS parameter~\cite{Ng:1993pv,Antusch:2021aiw,Soman:2024zor}, as illustrated in Fig.~\ref{fig:Expansion_EoS}. We denote this sequence of post-inflationary EoS parameters by $w_1,\, w_2,\, \ldots,\, w_n,\, w_{n+1}=1/3$, where $w_{n+1}$ is the EoS of the radiation-dominated hot Big Bang phase.

In addition, by assuming that the transition between any two successive epochs is sufficiently sharp, \textit{i.e.}, effectively instantaneous, one can employ the Israel junction conditions~\cite{Deruelle:1995kd} to relate the tensor mode functions across the transition. This procedure allows the Bogoliubov coefficients, and hence the mode functions in a succeeding epoch, to be determined directly from those in the preceding epoch. The junction conditions can be expressed as follows\,--
\begin{align}
    h_{k,\, b} ^{\lambda} (\tau_i) &= h_{k,\, a} ^{\lambda} (\tau_i) \,, ~~~~~~~~~\l[\text{Continuity}\r], \label{eq:continuity_hk}\\
    h_{k,\, b} ^{' \lambda} (\tau)  \Big|_{\tau = \tau_i} &= h_{k,\, a} ^{' \lambda} (\tau)  \Big|_{\tau = \tau_i}  \, , \quad \l[\text{Differentiability}\r] \label{eq:differentiability_hk}.
\end{align}
Here, $h_{k,\,b}^{\lambda}$ and $h_{k,\,a}^{\lambda}$ denote, respectively, the tensor mode functions immediately before and after a transition occurring at conformal time $\tau = \tau_i$. The general form of the tensor mode function in the $i^{\rm th}$ post-inflationary epoch (which ends at $\tau_i$), characterised by its corresponding Bogoliubov coefficients $A_{k,\,i}$ and $B_{k,\,i}$, can then be written as~(see Ref.~\cite{Soman:2024zor})

\begin{shaded}
\noindent{\bf \Brown Mode functions\,:}
\begin{equation}
    \boxed{~\bm{h_k ^{\lambda}} \l(y_{_k}\r) = \frac{1}{(\alpha_i \, y_{_k})^{\alpha_i - \half}} \l[\bm{A_{k, \, i}} \, J_{ \l(\alpha_i - \half \r)} (\alpha_i \, y_{_k}) +  \bm{B_{k , \, i}} \, J_{- \l(\alpha_i - \half \r)} (\alpha_i \, y_{_k}) \r]~} \label{eq:tensor_mode_ith_epoch} \; ,
\end{equation}
{\bf \Brown Bogoliubov coefficients\,:}
\begin{align}
    \bm{A_{k, \, i}} = \frac{\l( \alpha_i \, y_{_{k, \, i-1}} \r)^{\l(\alpha_i - \half \r)}}{\l( \alpha_{i-1} \, y_{_{k, \, i-1}} \r)^{\l(\alpha_{i-1} - \half \r)}}  \, \frac{\l[ \bm{A_{k, \, i-1}} \l( g_2 \, f_3 + g_4 \, f_1 \r) + \bm{B_{k, \, i-1}} \l( f_2 \, f_3 - f_4 \, f_1 \r) \r]}{f_1 \, g_3 + g_1 \, f_3} \, , \label{eq:coeff_A_k,n}\\
    \bm{B_{k, \, i}} = \frac{\l( \alpha_i \, y_{_{k, \, i-1}} \r)^{\l(\alpha_i - \half \r)}}{\l( \alpha_{i-1} \, y_{_{k, \, i-1}} \r)^{\l(\alpha_{i-1} - \half \r)}}  \, \frac{\l[ \bm{A_{k, \, i-1}} \l( g_2 \, g_3 - g_4 \, g_1 \r) + \bm{B_{k, \, i-1}} \l( f_2 \, g_3 + f_4 \, g_1 \r) \r]}{f_1 \, g_3 + g_1 \, f_3}\label{eq:coeff_B_k,n} \, ,
\end{align}
with {\bf \Brown dimensionless wavenumber\,:}
\beq
y_{_{k, \, i-1}} = \f{k}{a(\tau_{i-1}) H(\tau_{i-1})}\, ,
\label{eq:yk_ith_epoch_def}
\eeq
and the {\bf \Brown Bessel constants\,:}
\begin{equation}
\begin{aligned}
    g_1 &= J_{\l(\alpha_{i} - \half\r)} (\alpha_{i} \, y_{_{k, \, i-1}}) , &f_1 &= J_{-\l(\alpha_{i} - \half\r)} (\alpha_{i} \, y_{_{k, \, i-1}}) \, ,\\
    g_2 &= J_{\l(\alpha_{i-1} - \half\r)} (\alpha_{i-1} \, y_{_{k, \, i-1}}) , &f_2 &= J_{-\l(\alpha_{i-1} - \half\r)} (\alpha_{i-1} \, y_{_{k, \, i-1}}) \, , \\
    g_3 &= J_{\l(\alpha_{i} + \half\r)} (\alpha_{i} \, y_{_{k, \, i-1}}) , &f_3 &= J_{-\l(\alpha_{i} + \half\r)} (\alpha_{i} \, y_{_{k, \, i-1}}) \, ,\\
    g_4 &= J_{\l(\alpha_{i-1} + \half\r)} (\alpha_{i-1} \, y_{_{k, \, i-1}}) , &f_4 &= J_{-\l(\alpha_{i-1} + \half\r)} (\alpha_{i-1} \, y_{_{k, \, i-1}}) \,  \, .
\end{aligned}
\label{eq:Bessel_notations}
\end{equation}
\end{shaded}
It is important to emphasize that the analytical formalism employed in this work, based on the aforementioned usage of instantaneous transitions and Israel junction matching conditions, is reliable only when each post-inflationary epoch persists for a minimum duration of at least one $e$-fold. This requirement ensures that the background dynamics within each constant $w_i$ phase can be treated as quasi-stationary, thereby preserving the adiabatic evolution of tensor modes across transitions. If an epoch were to last for a substantially shorter duration, the assumption of a well-defined constant EoS would break down, and the discontinuous matching conditions would no longer provide a reliable description of the gravitational wave dynamics. In practice, this consistency condition has been incorporated within our numerical framework, \texttt{GWInSpect}, discussed in Sec.~\ref{sec:Num_GWInSpect}. 
\subsection{Spectral energy density and present-epoch frequency of inflationary GWs}
\label{sec:Omega_GWs_postinf}
The spectral energy density of the gravitational waves is a measure of the GW energy density per logarithm interval of frequency~\cite{Kohri:2018awv, Giovannini:2019ioo}, defined as,
\begin{equation}
    \Omega_{_{\text{GW}}}(\tau, k) =  \frac{1}{\rho_c}\, \frac{\d }{\d \ln k}\,\rho_{_{\rm GW}}(\tau, k) \, ,
    \label{eq:Omega_GW_def}
\end{equation}
where $\rho_{_{\rm GW}}$ is the energy density of GWs and $\rho_{c} (\tau) = 3 \, \mpl^2 \, H^2 (\tau)$ is the critical energy density. In the sub-Hubble regime\,--\,when a tensor mode is sufficiently inside the Hubble radius\,--\,expression for  the GW spectral energy density becomes~\cite{Soman:2024zor}
\begin{equation}\label{eq:Omega_GW_formula}
    \Omega_{_{\text{GW}}}(\tau, k)=\f{1}{12} \, \l[ \frac{k}{a(\tau) H(\tau) } \r]^2 \, {\cal P}_h(\tau,k) \, ,
\end{equation}
where ${\cal P}_h(\tau,k) = 2\,k^3\,|h_k|^2 /(2\pi^2)$ is the total tensor power spectrum (including both polarization states) at conformal time $\tau$. By evolving the tensor modes to the present epoch and considering the sub-Hubble regime, one can compute the tensor power spectrum today and thereby obtain the present-day spectral energy density of GWs, which, when averaged over the oscillations of individual modes,  takes the following form\,--
\begin{align}\label{eq:Omega_GW_formula_in_k}
    \overline{\Omega_{_{\rm GW}}(\tau_0, k)} = \l(\frac{\Omega_{\rm rad, \, 0} \, }{96 \pi^3}\r) \, \l(\frac{g_{*, \, {\rm r*}}}{g_{*, \, 0}} \r)\l(\frac{g_{s, \, 0}}{g_{s, \, {\rm r*}}}\r)^{4/3} \,    \frac{1}{y_{k, \, {\rm eq}}^2} \l[\tilde{A}^2_{k, \, {\rm MD}} +  \tilde{B}^2_{k, \, {\rm MD}}\r]  \,\frac{H^2_{\rm inf}}{\mpl^2} \l(\frac{k}{k_*}\r)^{n_{_T}} \, .
\end{align}
 Here, $g_*$ and $g_s$ denote, respectively, the effective number of relativistic degrees of freedom contributing to the energy and entropy densities of the Universe; $\Omega_{\rm rad,0}$ is the present-epoch radiation density parameter, $n_{_T}$ represents the primordial (inflationary) tensor spectral tilt, $k_*$ is the CMB pivot scale,  ${\rm r_*}$ denotes the beginning of hot Big Bang, and finally, $y_{_{k,{\rm eq}}} \equiv k/(a_{\rm eq} H_{\rm eq})$, with the subscript ``eq'' referring to the epoch of matter--radiation equality. The re-scaled Bogoliubov coefficients corresponding to the standard matter-dominated epoch are defined as $\tilde{A}_{k,{\rm MD}} \equiv A_{k,{\rm MD}}/h_{k,{\rm inf}}^{\lambda}$ and $\tilde{B}_{k,{\rm MD}} \equiv B_{k,{\rm MD}}/h_{k,{\rm inf}}^{\lambda}$. 

Eq.~\eqref{eq:Omega_GW_formula_in_k} can be conveniently re-expressed in terms of the present-day GW frequency, which is a directly measurable quantity in  GW observations. Since the primordial tensor spectral index is constrained to be very small ($|n_{_T}| \ll 1$), its effect on the overall amplitude and tilt of the spectrum is negligible compared to that arising from the post-inflationary dynamics. Therefore, we set $n_{_T} = 0$ throughout this analysis in order to isolate and highlight the imprints of the pre--hot Big Bang expansion history on the morphology of the GW spectral energy density. The resulting expression for the present-day spectral energy density of inflationary GWs, written as a function of the GW frequency, takes the following form~\cite{Soman:2024zor}\,--
\begin{shaded}
\noindent{\bf \Brown Spectral energy density\,:}
    \begin{align}
    \label{eq:Omega_GW_fun_of_freq}
    \boxed{~h^2 \, \bm{\Og ^{(0)}} (f_k) = \frac{1}{96 \pi^3} \,\frac{g_{*, \, {\rm r*}}}{g_{*, \, 0}} \l(\frac{g_{s, \, 0}}{g_{s, \, {\rm r*}}}\r)^{4/3} \, h^2 \Omega_{\rm rad, \, 0} \,    \l(\frac{f_{\rm eq}}{f_k} \r)^2 \l[\bm{\tilde{A}^2_{k, \, {\rm MD}}} +  \bm{\tilde{B}^2_{k, \, {\rm MD}}}\r]  \l(\frac{H_{\rm inf}}{\mpl}\r)^2 ~}\, ,
    \end{align}
    where the {\bf \Brown rescaled Bogoliubov coefficients} are
    \beq
\bm{\tilde{A}_{k,{\rm MD}}} = \f{A_{k,{\rm MD}}}{h_{k,{\rm inf}}^{\lambda}}\, , \quad \bm{\tilde{B}_{k,{\rm MD}}} = \f{B_{k,{\rm MD}}}{h_{k,{\rm inf}}^{\lambda}} \, .
\label{eq:Bog_A_B_rescaled}
    \eeq
Relation between the present-day frequency ($f_k$) of a tensor mode to the \textit{effective} temperature ($T_k$), and  energy scale ($E_k$), of the universe at the Hubble-entry time of that mode is given by\,--
\medskip

\noindent{\bf\Brown Present-day Frequency of GWs\,:}
\begin{align}\label{eq:rel_bet_freq_and_temp}
    \boxed{\bm{\frac{f_k}{{\rm Hz}}} = 7.43 \t 10^{-8} \l(\frac{g_{s, \, 0}}{g_{s,T_k}} \r)^{1/3} \l(\frac{g_{*,T_k}}{90} \r)^{1/2}  \l(\bm{\frac{T_k}{ {\rm GeV}}} \r) = 1.03 \t 10^{-8} \l(\frac{g_{s, \, 0}}{g_{s,T_k}} \r)^{1/3}  g_{*,T_k} ^{1/4}  \l(\bm{\frac{E_k}{ {\rm GeV}}} \r)} \, .
\end{align}
The  spectral tilt of $\Og ^{(0)}(f)$ for the range of modes making an Hubble-entry during the $i^{\rm th}$ post-inflationary epoch depends on the EoS of the epoch in the following way\,--
\medskip

\noindent{\bf \Brown Spectral tilt of GWs\,:}
\begin{equation}
   \boxed{~ \bm{n_{_{\rm GW}}^{(i)}} \equiv \frac{\d \ln \Og ^{(0)} (f_k)}{\d \ln f_k} =  \bm{2 \l( \frac{w_i - 1/3}{w_i + 1/3} \r)} ~} \,, \quad \text{for} \quad \bm{f_{\{i\}} < f_k < f_{\{i-1\}}} \, .
    \label{eq:spectral_tilt}
\end{equation}
\vspace{-0.1in}
\end{shaded}
\subsection{Analytical treatment of the BBN bound on stochastic GWs}
\label{sec:BBN_bound_Anlyt}
Observations of the primordial light-element abundances from BBN impose a stringent upper bound on the total energy density of primordial GWs, given by
\begin{align}\label{eq:BBN_constraint_integral}
    h^2 \int_{f_{\rm BBN}}^{f_{\rm end}} \d \ln f ~ \Og(\tau_0, f) \;<\; 1.13 \times 10^{-6} \, ,
\end{align}
where the integration extends from the frequency $f_{\rm BBN}$ corresponding to the tensor mode that re-entered the Hubble radius at the onset of BBN, up to $f_{\rm end}$ associated with the mode that re-entered at the end of inflation.  
In practice, since within any given $i^{\rm th}$ post-inflationary epoch the GW spectral energy density scales as $\Omega_{\rm GW} \propto f^{\,n_{_{\rm GW}}^{(i)}}$ (with $n_{_{\rm GW}}^{(i)}$ defined in Eq.~\eqref{eq:spectral_tilt}), the BBN integral in Eq.~\eqref{eq:BBN_constraint_integral} can be evaluated efficiently by performing a piecewise analytical integration across the different epochs, as derived in Ref.~\cite{Soman:2024zor}.  
This leads to the following compact condition ensuring consistency with the BBN bound\,--

\begin{shaded}
\noindent{\bf \Brown BBN Constraint\,:}
\ber
     \Biggl[  \ln{\l( \frac{f_{\rm r*}}{f_{\rm BBN}} \r)}  &+&  
       \frac{{\cal F}_n}{2(1-\alpha_n)}
          \l\{ \l(\frac{f_{n-1}}{f_n} \r)^{2(1-\alpha_n)} -1 \r\} 
        + \frac{{\cal F}_{n-1}}{2(1-\alpha_{n-1})}
          \l\{ \l(\frac{f_{n-2}}{f_{n-1}} \r)^{2(1-\alpha_{n-1})} -1 \r\}  \nonumber \\
    && \hspace{-0.2in} + ..... +  \frac{{\cal F}_1}{2(1-\alpha_1)}
          \l\{ \l(\frac{f_{\rm end}}{f_{1}} \r)^{2(1-\alpha_{1})} -1 \r\} 
    \Biggr] ~< ~ 1.13 \t 10^{-6} \t \l( h^2 \Omega_{\rm GW}^{0, \, \rm RD} \r)^{-1} \, , \label{eq:BBN_Bound_Piecewise_Approx} 
\eer
where  each $f_i$ corresponds to the frequency of a tensor mode that re-entered the Hubble radius at the end of $i^{\rm th}$ epoch and while ${\cal F}_i$ is defined as
\begin{equation}
    {\cal F}_i = \prod_{m = i}^n \l( \frac{f_m}{f_{m+1}} \r)^{2(1-\alpha_{m+1})} \label{eq; coeff F_i} \,, \quad \text{and} \quad i \in \{1, 2, ... , n \},
\end{equation}
and 
\begin{equation}
     \Omega_{\rm GW}^{0, \rm RD} \simeq \frac{1}{24} \, \Omega_{\rm rad,\, 0} \times \f{2}{\pi^2} \times \l(\f{H_{\rm inf}}{m_p}\r)^2
     \label{eq:value_Omega_GW^rad} \, .
\end{equation}
\vspace{-0.15in}
\end{shaded}
\vspace{-0.15in}
\section{The morphological zoo of inflationary gravitational wave spectra}
\label{sec:Inf_GW_Zoo}
In this section, we undertake a detailed investigation into the diverse shapes that the inflationary GW spectrum can exhibit as a consequence of different post-inflationary (pre-BBN) expansion histories of the Universe. As emphasised above, the evolution of tensor modes across successive epochs, each characterized by a distinct EoS parameter $w_i$, imprints a rich structure on the spectral energy density $\Omega^{(0)}_{\rm GW}(f)$. Depending on the sequence, duration, and relative energy scales of these epochs, the resulting spectra can display a wide range of morphologies---including monotonic (blue- or red-tilted), convex, concave, as well as non-monotonic or multi-peaked behaviours. 

Our goal here is to systematically map this {\em zoo of inflationary GW spectra} in a model-agnostic manner, relying solely on the macroscopic post-inflationary dynamics encapsulated by the EoS parameters $\{w_i\}$ and the transition scales $\{E_i\}$. By varying these parameters within the framework established in Sec.~\ref{sec:Omega_GW}, we illustrate how even simple sequences of instantaneous transitions can give rise to remarkably distinct spectral features. Such an analysis not only highlights the sensitivity of $\Omega^{(0)}_{\rm GW}(f)$ to the early expansion history, but also provides valuable intuition for interpreting future detections of a stochastic GW background in terms of the underlying kinematics of cosmic evolution.

\begin{figure}[htb] 
    \centering
    \includegraphics[width = 0.8\textwidth]{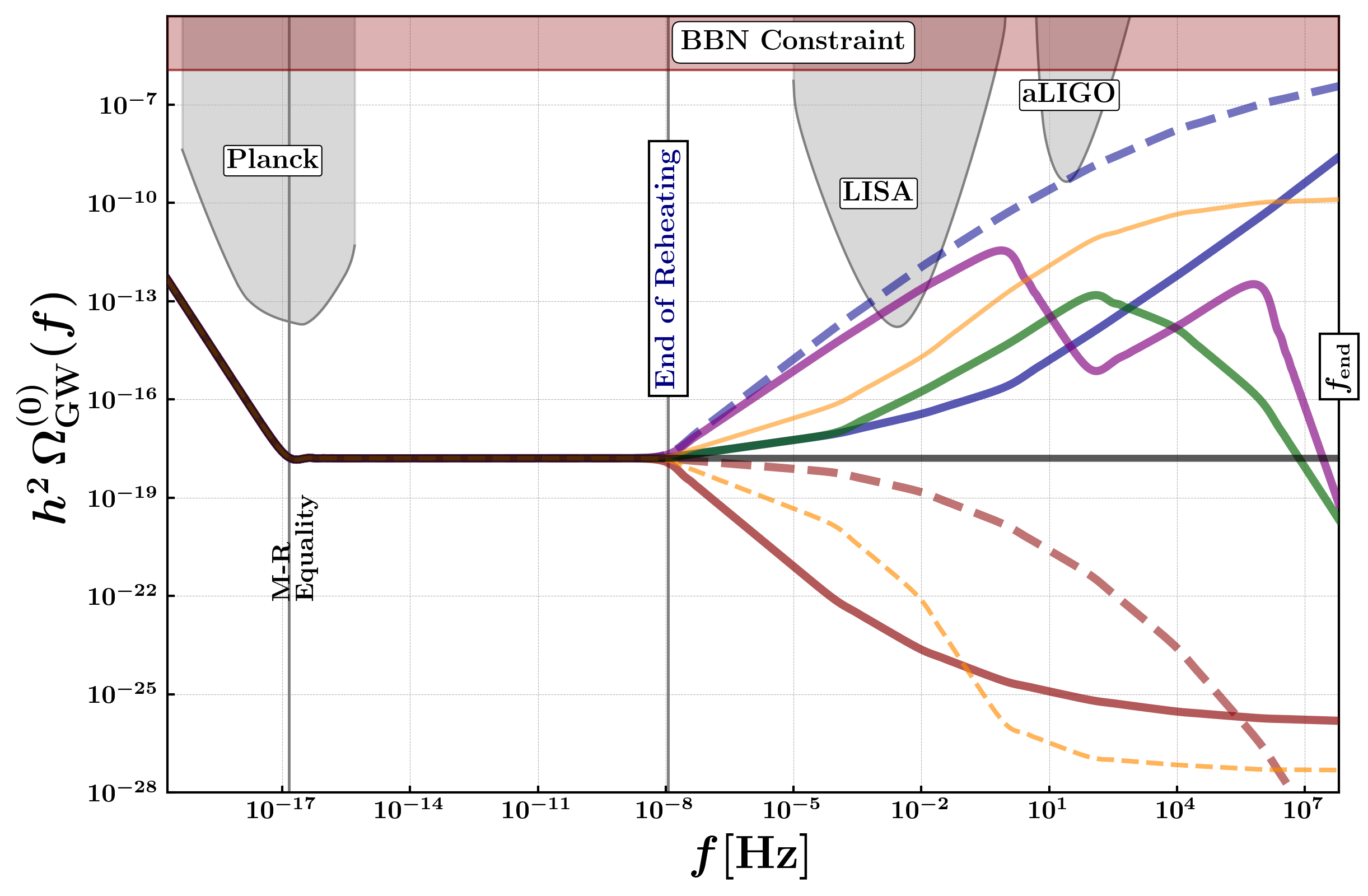}
    \caption{Schematic illustration of representative morphologies of the present-day spectral energy density of inflationary gravitational waves, $\Omega^{(0)}_{\rm GW}(f)$, as a function of frequency $f$ (for various sequences of 7 pre-hot Big Bang epochs after inflation). The solid (dashed) {\blue \bf blue curve} represents a convex (concave), monotonically increasing  spectrum arising from a decreasing (increasing)  sequence of post-inflationary EoS parameters $w_1 > w_2 > \cdots > w_n $~(for concave the sequence~$w_1 < w_2 < \cdots < w_n $ ), corresponding to an overall blue-tilted spectrum.
    Similarly the solid (dashed) {\Brown \bf brown curve} denotes a convex (concave), monotonically decreasing spectrum produced by a decreasing (increasing) sequence of EoS parameters~$w_1 > w_2 > \cdots > w_{n}$~(for concave shape~$w_1 < w_2 < \cdots < w_n$), resulting in a red-tilted spectrum. Non-monotonic spectra are illustrated by the  {\fgreen \bf green curve} for a single peak  and the {\violet \bf purple curve} for two peaks (and a dip). The solid (dashed) {\bf \borange orange curve} corresponds to monotonically increasing (decreasing) spectrum which is neither convex nor concave.  The solid  {\bf black curve} illustrates an approximately scale-invariant spectrum corresponding to a radiation-like EoS $w=1/3$.
    The grey-shaded regions indicate the approximate sensitivities of space-based detectors (\texttt{LISA}) and ground-based interferometers (\texttt{aLIGO}), while the red-shaded region marks the BBN constraint. The figure serves to highlight the broad variety of inflationary GW spectral shapes possible for different post-inflationary expansion histories.}
    \label{fig:Omega_GW_Morphology_Schematic}
\end{figure}

To be precise, different shapes of the GW spectra correspond to the following sequence of the post-inflationary  expansion history\,--
\vspace{-0.1in}
\begin{shaded}
\vspace{-0.15in}
\begin{enumerate}
\item The spectrum is monotonic when either $w_i > 1/3$ (blue-tilted) or $w_i < 1/3$ (red-tilted) for all pre-hot Big Bang epochs $i$.  
\item A monotonic spectrum is convex when $w_1 > w_2 > w_3 \cdots > w_n$, while it is concave when  $w_1 < w_2 < w_3 \cdots < w_n$. In the absence of a strict ordering among the EoS parameters, the resulting monotonic spectrum ceases to exhibit a definite curvature and becomes neither convex nor concave.
\item Non-monotonic behaviour with peaks and/or troughs appear when some of the EoS parameters obey $w_i > 1/3$, while others obey $w_j < 1/3$.
\end{enumerate}
\vspace{-0.2in}
\end{shaded}    
\vspace{-0.1in}
A schematic illustration of representative spectral morphologies is shown in Fig.~\ref{fig:Omega_GW_Morphology_Schematic}, where the spectral energy density $\Omega^{(0)}_{\rm GW}(f)$ is plotted as a function of the present-day GW frequency $f$. The grey-shaded regions indicate approximate detector sensitivity regions (\texttt{Planck} CMB, \texttt{LISA}, \texttt{aLIGO}) and the horizontal dashed line (with reddish shade) schematically marks the BBN constraint. We do not display sensitivity regions of other GW detectors, which were displayed in Ref.~\cite{Soman:2024zor}. We adhere to the following conventions in all our GW spectral energy density plots\,--
\vspace{-0.1in}
\begin{shaded}
\vspace{-0.2in}
\begin{enumerate}
    \item The present-day spectral energy density of inflationary GWs, $\Omega^{(0)}_{\rm GW}(f)$, is plotted as a function of the present-day frequency $f$, with both axes in logarithmic scale, as is customary in the literature.    
    \item The tensor-to-scalar ratio during inflation is fixed at $r = 0.001$, which corresponds to an inflationary energy scale of 
    $E_{\rm inf} \simeq 5.8 \times 10^{15}\,{\rm GeV}$ 
    (see App.~A.2 of Ref.~\cite{Soman:2024zor}).    
    \item The present-day frequency of the tensor mode that re-entered the Hubble radius at the end of inflation can be obtained using Eq.~\eqref{eq:rel_bet_freq_and_temp} as 
    $f_{\rm end} \simeq 6.4\times 10^7\,{\rm Hz}$. 
    We consider this value as the absolute upper limit (UV cut-off) on the GW frequency in all our plots.   
    \noindent The lower frequency limit (IR cut-off) is set at 
    $f_{\rm IR} \simeq 2\times 10^{-20}\,{\rm Hz}$, 
    corresponding to modes that re-enter the Hubble radius just prior to the onset of the dark energy–dominated accelerated expansion of the Universe.    
    \item Frequencies corresponding to the transition between successive cosmic epochs are indicated by thin vertical grey lines in all the plots.
\end{enumerate}
\vspace{-0.1in}
\end{shaded}
\vspace{-0.1in}
\subsection{Scenarios with monotonic spectra of inflationary gravitational waves}
\label{sec:Inf_GW_Zoo_monotonic}

Monotonic spectra of inflationary GWs arise when the sequence of post-inflationary equation-of-state parameters, starting from the first epoch after inflation, evolves to the final epoch prior to the hot Big Bang phase, by remaining either  stiffer ($w_i>1/3$) or  softer ($w_i<1/3$) than a radiation-like EoS $w=1/3$. A sequence of stiffer-than-radiation EoS parameters leads to blue-tilted GW spectrum, while a softer-than-radiation sequence leads to red-tilted GW spectrum, as mentioned before. The commonly assumed scenario of a single constant EoS during the entire post-inflationary evolution is a particular case within this category. 

An \emph{increasing} sequence of EoS parameters, \textit{i.e.}, $w_1 < w_2 < \cdots < w_n$, results in a \emph{concave}-shaped monotonic spectrum of GWs. Conversely, a \emph{decreasing} sequence, $w_1 > w_2 > \cdots > w_n$,  leads to a progressively softer background and produces a \emph{convex}-shaped spectrum. Of particular significance are the blue-tilted spectra of GWs (which may be convex or concave or neither)  that arise,  when $w \geq 1/3$.  From a phenomenological standpoint, as it enhances the GW amplitude at higher frequencies, potentially bringing the signal within the sensitivity range of current and upcoming detectors~\cite{Sahni:1990tx,Caprini:2018mtu,Figueroa:2019paj}, as also discussed extensively in our previous work~\cite{Soman:2024zor}.

\begin{figure}[hbt]
    \centering
    {\includegraphics[width=0.485\textwidth]{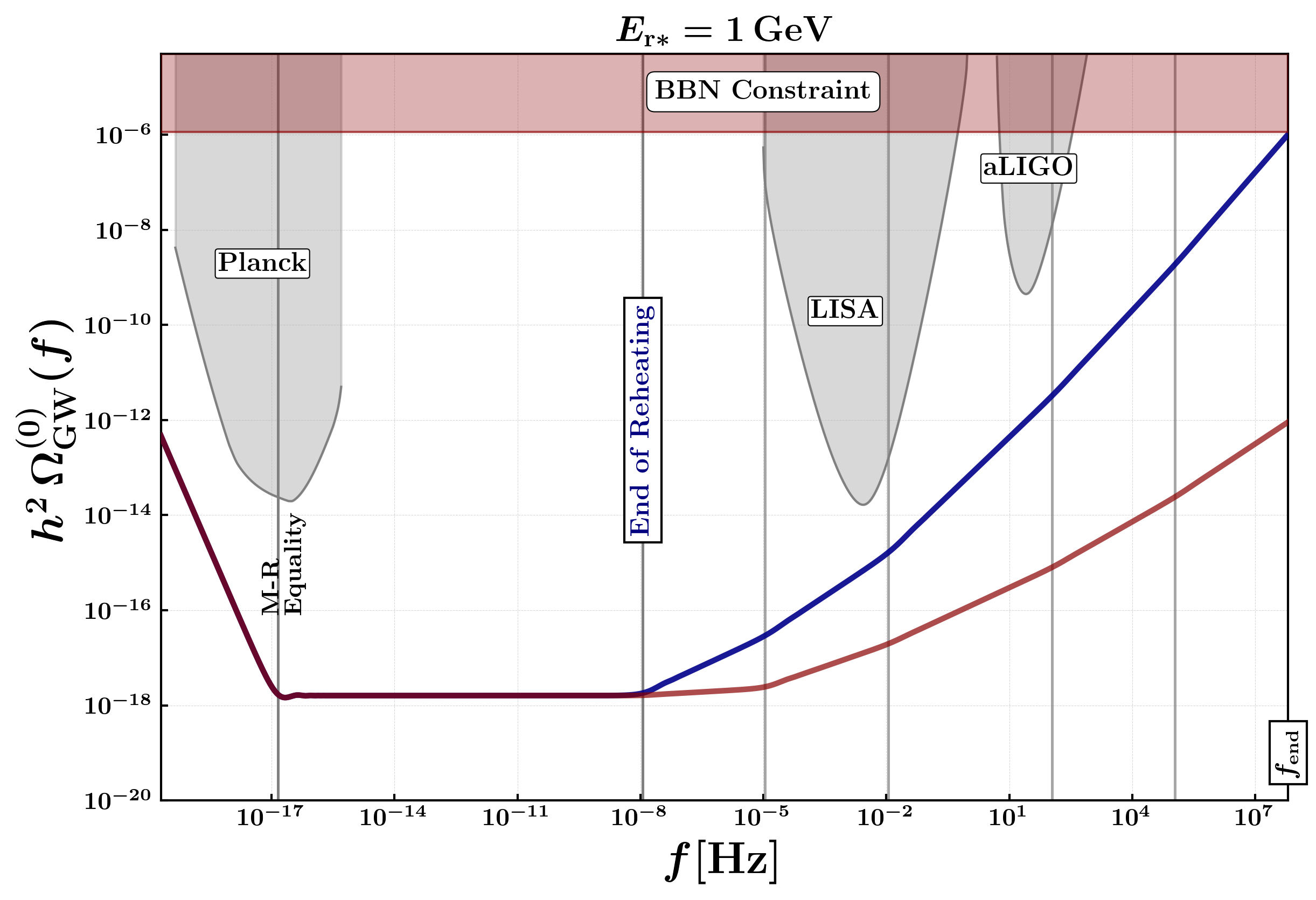}}
    {\includegraphics[width=0.485\textwidth]{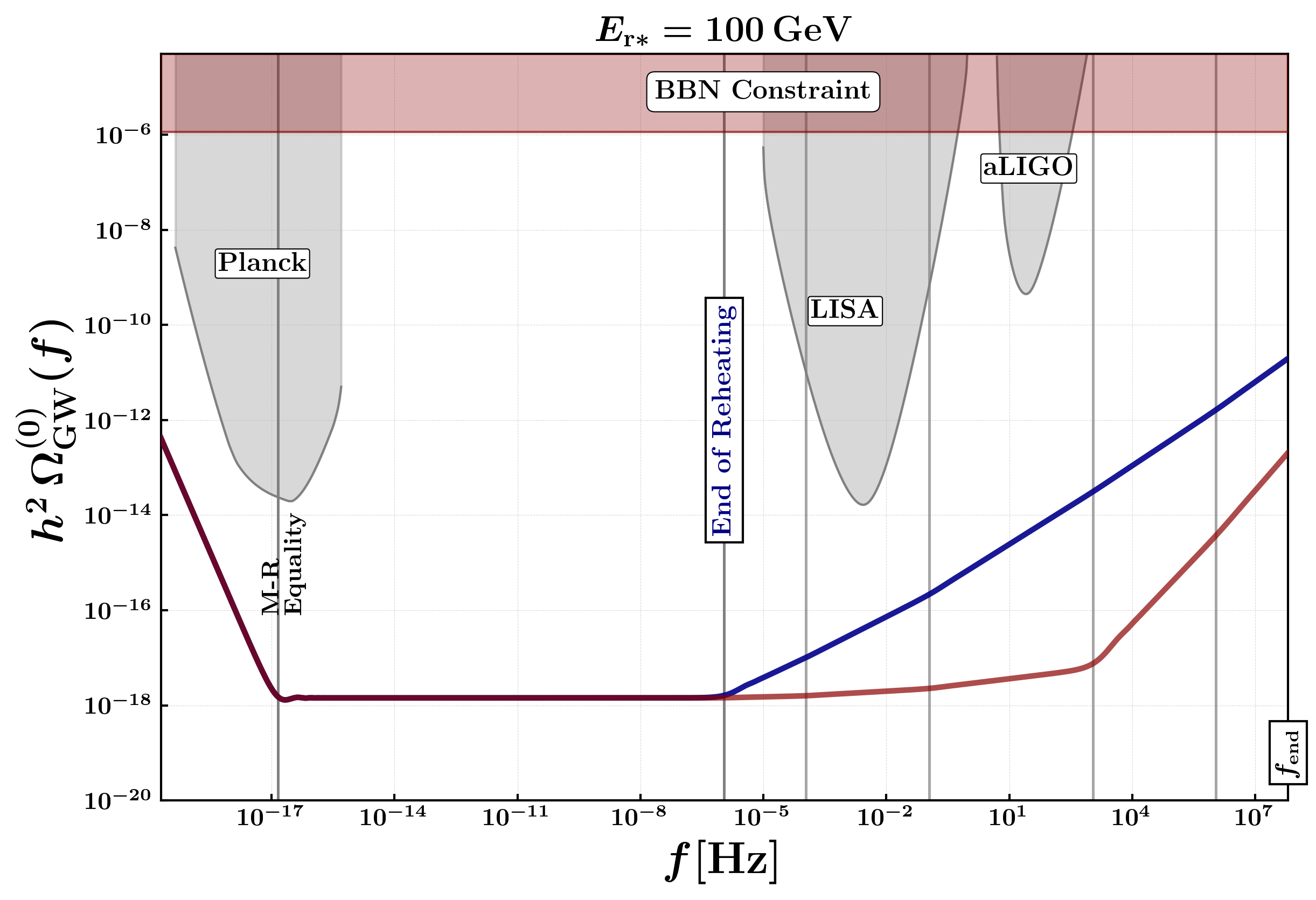}}
    {\includegraphics[width=0.485\textwidth]{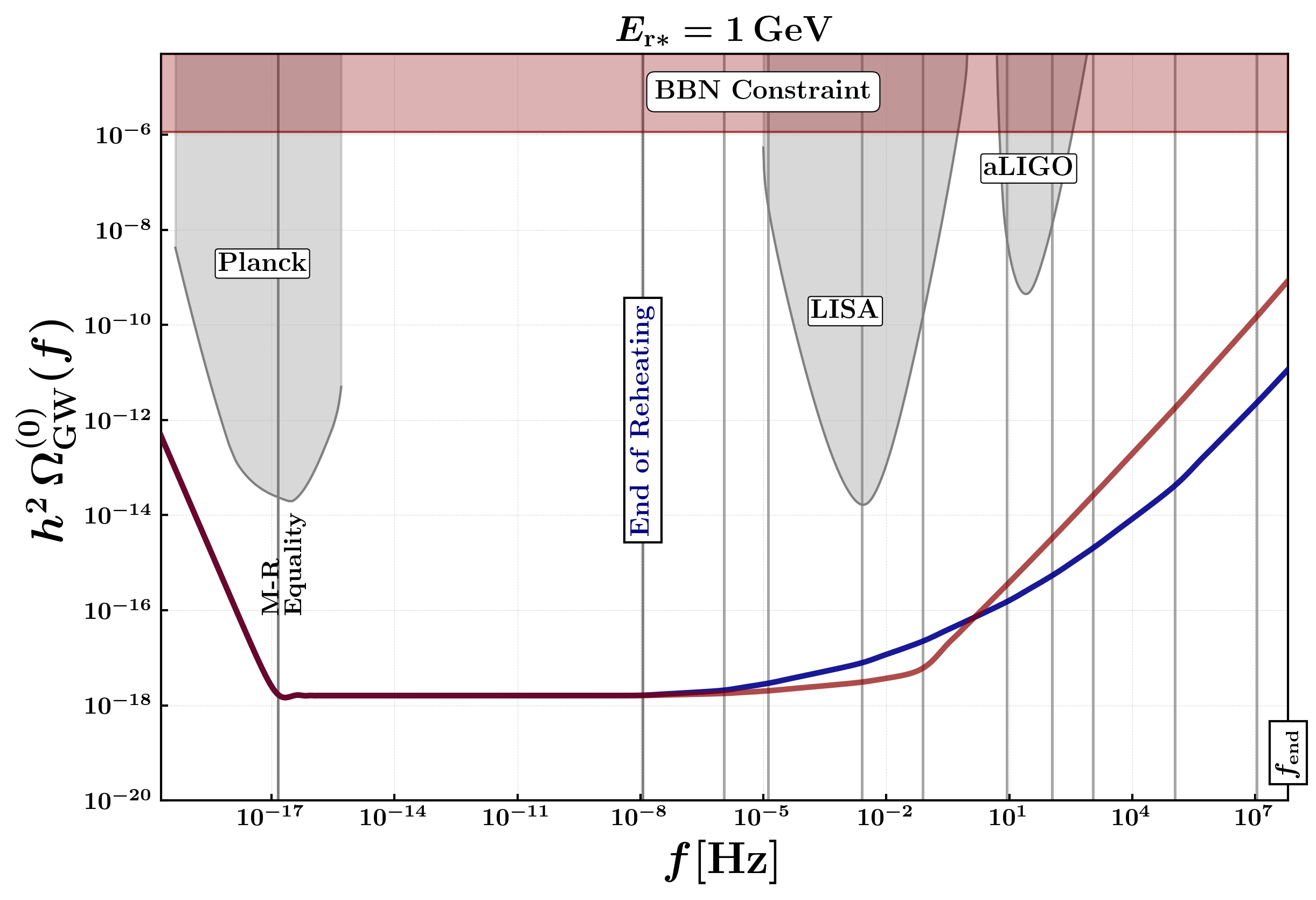}}
    {\includegraphics[width=0.485\textwidth]{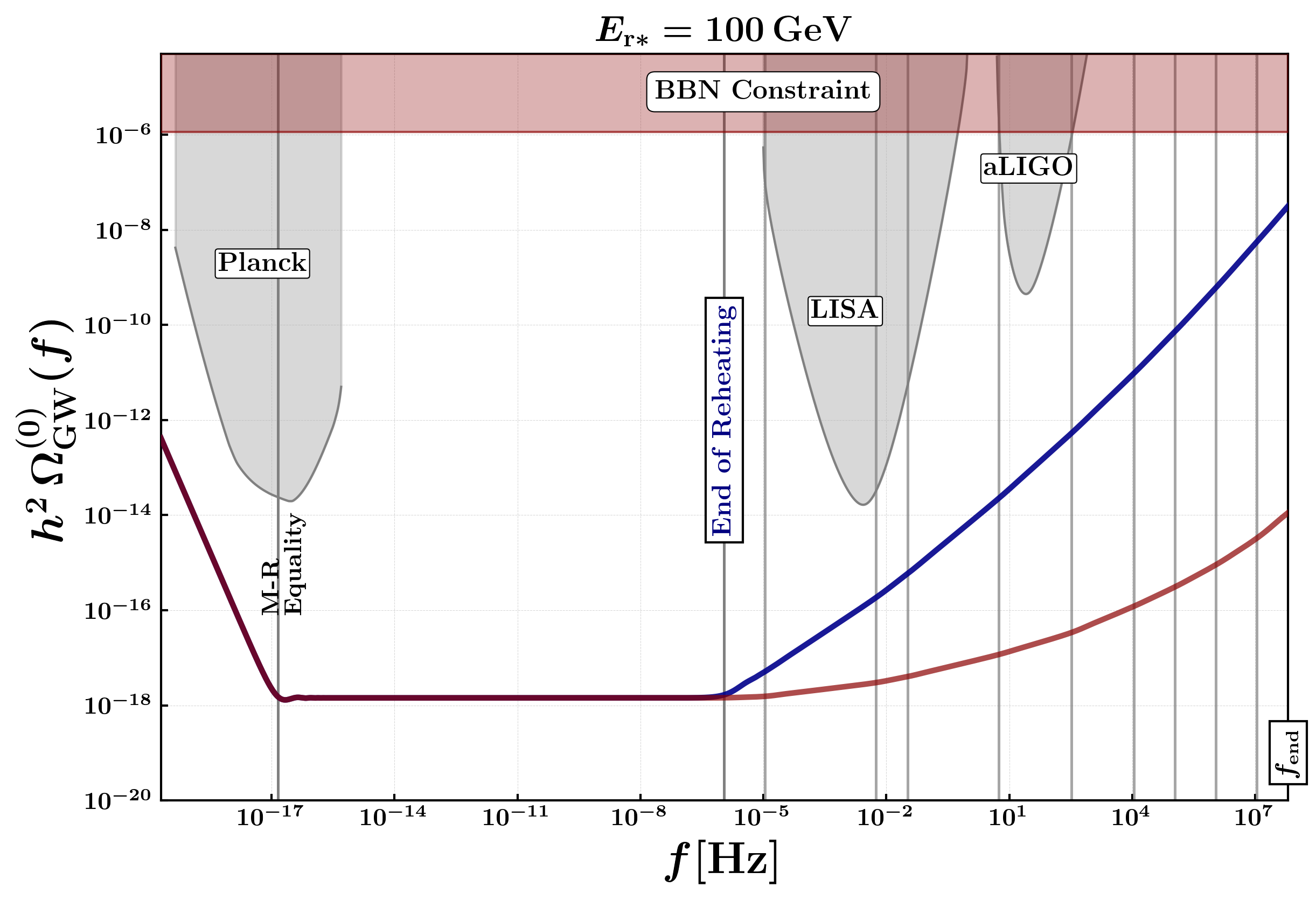}}
    \caption{Spectral energy density of inflationary GWs, $\Omega^{(0)}_{\rm GW}(f)$, for the case of monotonic blue-tilted, {\bf convex spectra} generated by strictly {\em decreasing sequences of post-inflationary EoS} parameters $\l(w_1>w_2>\cdots>w_n>w_{\rm rad}=1/3\r)$. The {\bf top row} contains 5 different post-inflationary epochs, while the {\bf bottom row} contains 10 epochs. The {\bf left column} corresponds to an energy scale of $E_{\rm r*}=1 \, \mathrm{GeV}$ at the beginning of hot Big Bang, while the {\bf right column} corresponds to $E_{\rm r*}=100 \, \mathrm{GeV}$. Transitions between successive epochs are marked by grey vertical lines. The BBN constraint is illustrated by the red-shaded region, and the sensitivity curves for \texttt{LISA}, \texttt{Planck}, and \texttt{aLIGO} are displayed by grey-shaded regions. [None of the curves violate the existing constraints from BBN and \texttt{aLIGO}.]}
    \label{fig:convex_monotonic}
\end{figure}

\subsubsection{Blue-tilted inflationary GWs with convex-shaped  spectra}
\label{sec:Inf_GW_Zoo_monotonic_convex}
A convex-shaped, monotonically increasing spectrum of inflationary GWs arises when the post-inflationary epochs are characterized by a sequence of progressively softer equations of state, $w_1 > w_2 > w_3 > \ldots > w_{\rm rad} = 1/3$. In such scenarios, the spectral energy density $\Omega^{(0)}_{\rm GW}(f)$ grows with frequency ($n_{_{\rm GW}}(f) > 0$), and the rate of growth also gradually increases, \textit{i.e.}, $\l[\d/(\d \ln f)\r] n_{_{\rm GW}}(f) > 0$, producing a convex curvature in the $\log\Omega_{\rm GW}$\,---\,$\log f$ plane. This behaviour typically originates from the diminishing amplification of the tensor modes as the expansion history transitions from a stiffer to a softer EoS regime. The highest blue tilt results from the first post-inflationary epoch with the stiffest EoS, while the convexity encodes the hierarchical duration and energy scales of the subsequent, softer epochs. A representative class of convex-shaped blue-tilted GW spectra is shown in~Fig.~\ref{fig:convex_monotonic}.

Such convex monotonic spectra are particularly relevant for phenomenological studies, since they can feature a gradual spectral saturation within the frequency bands probed by future space-based interferometers such as \texttt{LISA}, \texttt{BBO}, and \texttt{DECIGO}, with the caveat that a substantial growth of $\Omega^{(0)}_{\rm GW}(f)$ at higher frequencies might lead to a violation of the BBN constraint~\eqref{eq:BBN_Bound_Piecewise_Approx}.

\begin{figure}[hbt]
    \centering
    {\includegraphics[width=0.485\textwidth]{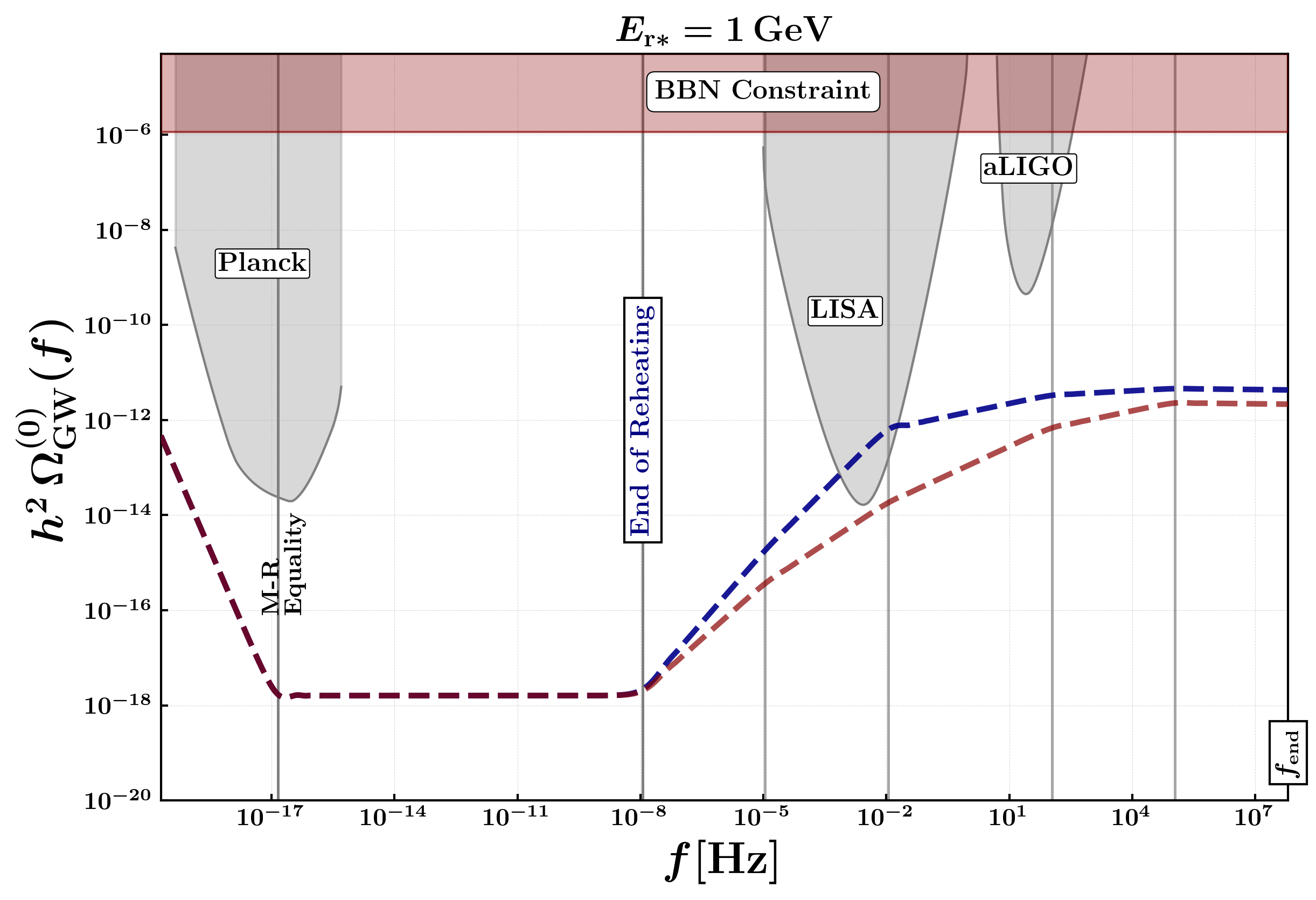}}
    {\includegraphics[width=0.485\textwidth]{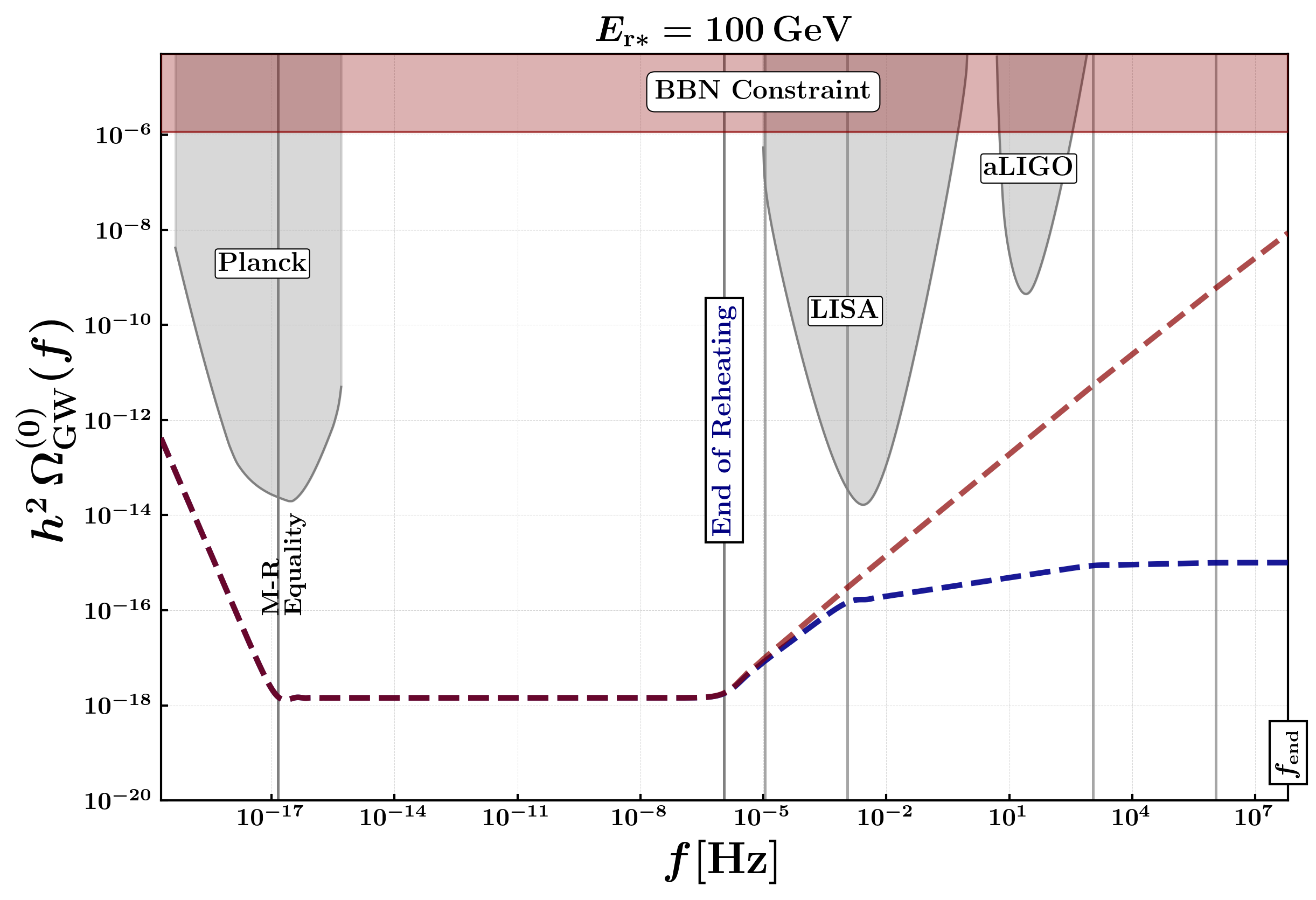}}
    {\includegraphics[width=0.485\textwidth]{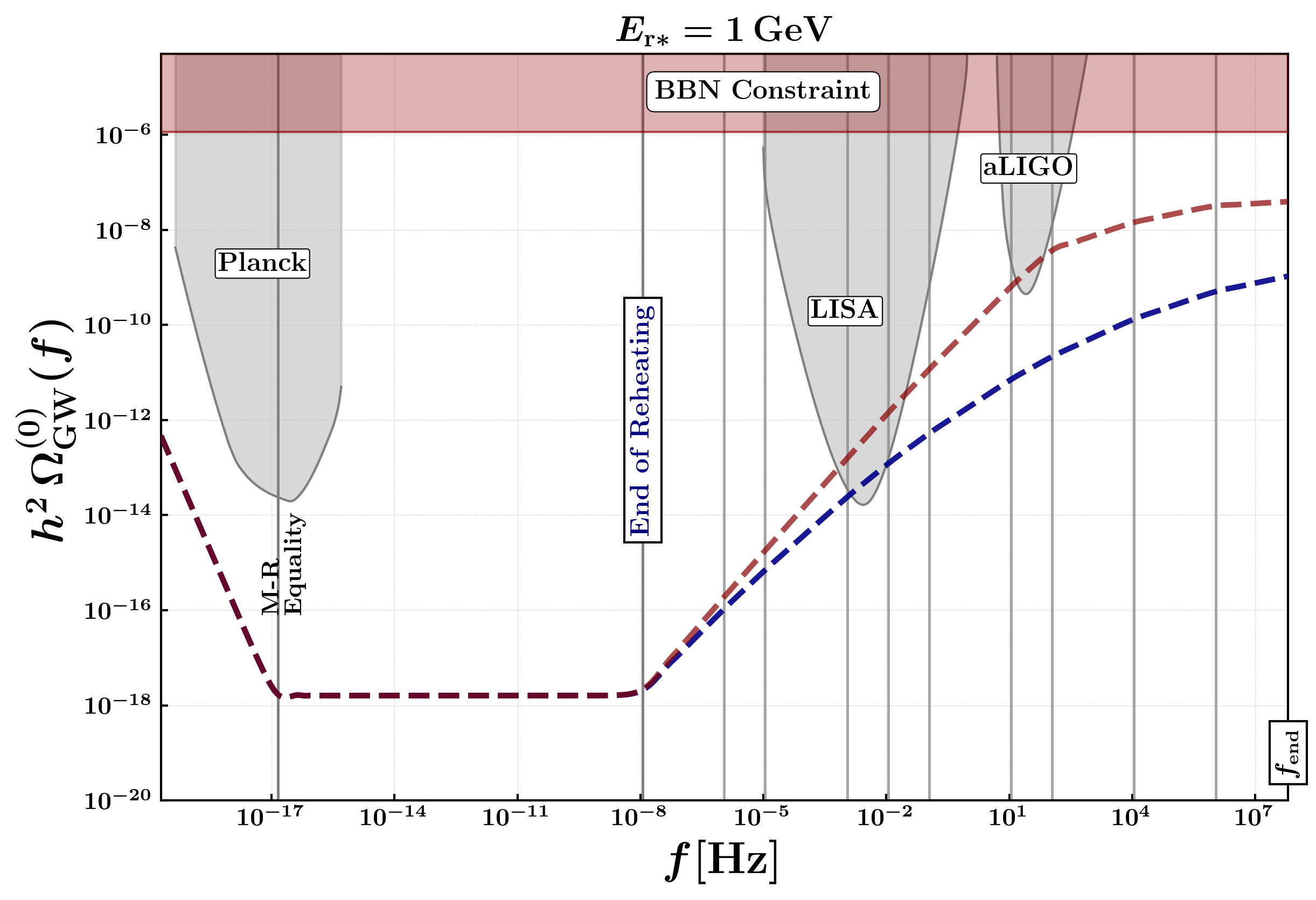}}
    {\includegraphics[width=0.485\textwidth]{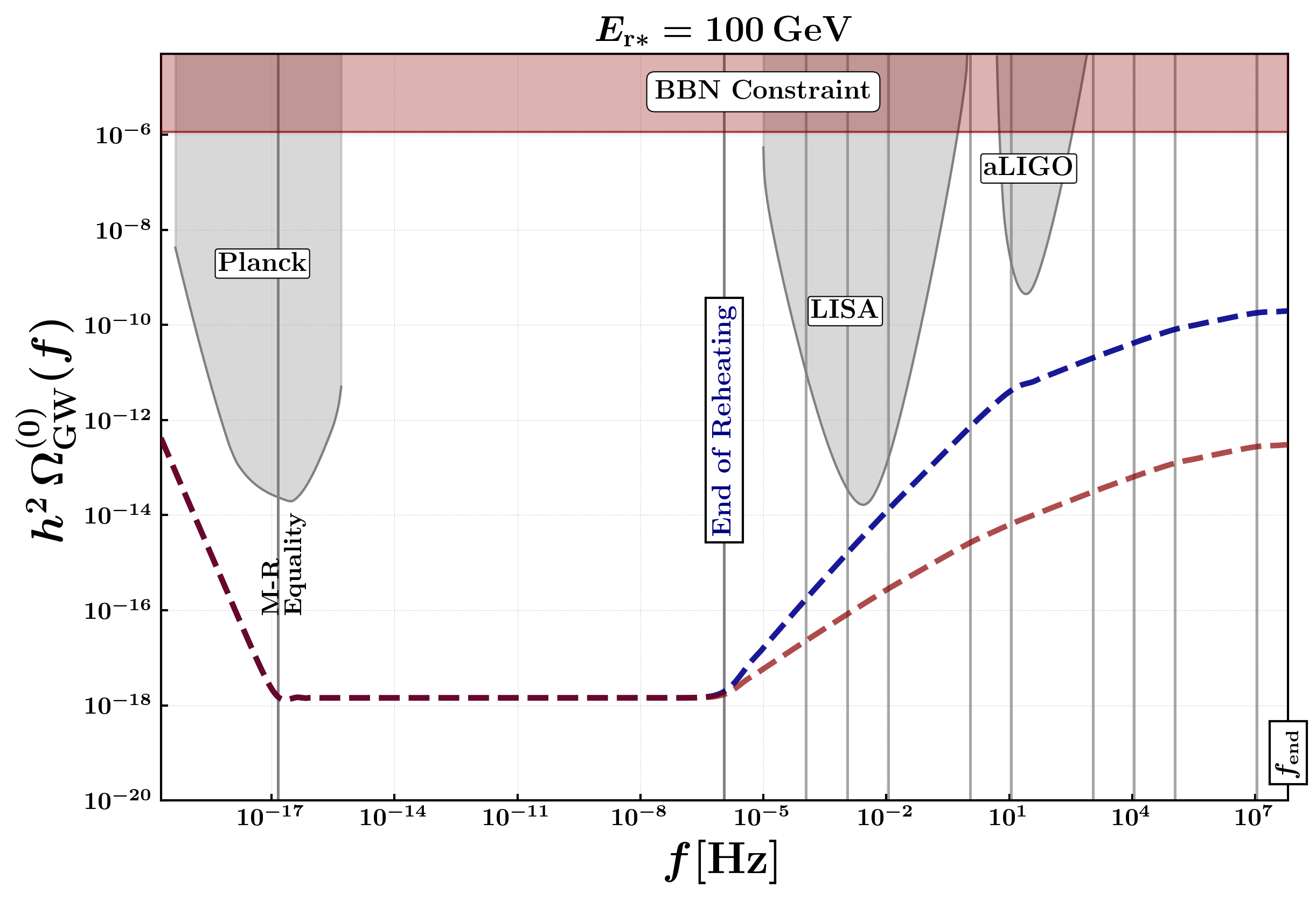}}
    \caption{
        Spectral energy density of inflationary GWs, $\Omega^{(0)}_{\rm GW}(f)$, for the case of monotonic blue-tilted, {\bf concave spectra} generated by  strictly {\em increasing sequences of post-inflationary EoS}  parameters $\l(\bm{w_1 < w_2 < \cdots < w_n} \r)$. The {\bf top row} contains 5 different post-inflationary epochs, while the {\bf bottom row} contains 10 epochs. The {\bf left column} corresponds to an energy scale of $E_{\rm r*}=1 \, \mathrm{GeV}$ at the beginning of hot Big Bang, while the {\bf right column} corresponds to $E_{\rm r*}=100 \, \mathrm{GeV}$. Transitions between successive epochs are marked by grey vertical lines. The BBN constraint is illustrated by the red-shaded region, and the sensitivity curves for \texttt{LISA}, \texttt{Planck}, and \texttt{aLIGO} are displayed by grey-shaded regions. [None of the curves violate the existing constraints from BBN and \texttt{aLIGO}, except for the red-dashed curve in the bottom-left panel.]
        }
    \label{fig:concave_monotonic}
\end{figure}

\subsubsection{Blue-tilted inflationary GWs with concave-shaped spectra}
\label{sec:Inf_GW_Zoo_monotonic_concave}
A concave-shaped, monotonically increasing spectrum of inflationary GWs typically results from a post-inflationary expansion history governed by a sequence of increasingly stiffer equations of state, $w_1 < w_2 < w_3 < \ldots < w_n$. In this case, the spectral energy density $\Omega^{(0)}_{\rm GW}(f)$ increases with frequency, exhibiting an overall blue tilt ($n_{_{\rm GW}}(f) > 0$). However, slope of the spectrum gradually decreases, \textit{i.e.}, $\l[\d/(\d \ln f)\r] n_{_{\rm GW}}(f) < 0$, leading to a concave curvature in the $\log\Omega_{\rm GW}$\,---\,$\log f$ plane.  A representative class of concave-shaped blue-tilted GW spectra is shown in~Fig.~\ref{fig:concave_monotonic}. Such a spectral shape  encodes the gradual hardening of the background EoS\,---\,undergoing an ordered  sequence of  transitions from softer to stiffer dynamics prior to the onset of radiation domination. They also possess a clear phenomenological advantage\,: their concave shape enables the spectrum to enter the sensitivity bands of space-based (and forthcoming ground-based) detectors at intermediate frequencies, while remaining consistent with the BBN bound due to the gradual flattening of $\Omega^{(0)}_{\rm GW}(f)$ at higher frequencies.
 
\subsection{Scenarios with non-monotonic spectra of inflationary gravitational waves}
\label{sec:Inf_GW_Zoo_nonmonotonic}

In the previous subsection, we focused on monotonically increasing classes of inflationary GW spectra arising from post-inflationary expansion histories characterised by an ordered\,--\,either steadily increasing or decreasing\,--\,sequence of EoS parameters. In contrast, in the absence of such an ordered sequence of $\lbrace w_i\rbrace$, we generically expect non-monotonic features in $\log\Omega^{(0)}_{\rm GW}(f)$\,---\,$\log f$ plane.

Such scenarios typically occur when the post-inflationary universe experiences alternating epochs of softer and stiffer equations of state prior to the commencement of the hot Big Bang phase. A gradual transition from softer/stiffer epochs to  stiffer/softer epochs  imprints a local feature\,--\,such as a peak (bump) or a trough (dip)\,--\,in the spectral slope of $\Omega^{(0)}_{\rm GW}(f)$. The resulting spectra can thus exhibit multiple turning points, corresponding to characteristic frequencies at which the horizon re-entry of tensor modes coincides with the transition between successive EoS phases. This stands in contrast to the conventional expectation in the literature that inflationary gravitational waves yield strictly monotonic spectra.

In this subsection, we illustrate representative examples of non-monotonic spectra of inflationary GWs, highlighting the variety of shapes that may arise from different sequences of post-inflationary EoS parameters. For concreteness, we restrict our analysis to spectra exhibiting up to two distinct peaks in $\Omega^{(0)}_{\rm GW}(f)$, as these already encompass the essential phenomenology associated with multiple transitions in the equation of state. However, the underlying formalism developed in our previous work~\cite{Soman:2024zor} and employed here is fully general, and allows the user to construct GW spectra with an arbitrary number of transitions and spectral features\,\footnote{with the caveat that each epoch must last for at least one $e$-fold in order for our analytical approach to remain valid, as discussed in Secs.~\ref{subsec:tensor_postinf}~and~\ref{sec:GWInSpect_Inputs}.}. The shapes discussed below therefore serve as templates for understanding how alternating soft and stiff epochs in the post-inflationary history can lead to a rich variety of  GW morphologies.

\subsubsection{Non-monotonic spectra for a soft EoS immediately after inflation}
\label{sec:Inf_GW_Zoo_nonmonotonic_soft}

A non-monotonic spectrum can arise when the post-inflationary universe begins with a relatively soft EoS ($w_1 < 1/3$) before transitioning to stiffer or radiation-like epochs. In this case, the early suppression of tensor modes re-entering during the soft phase results in an initial red tilt of $\Omega^{(0)}_{\rm GW}(f)$ at high frequencies. Subsequent transitions to larger  values of $w_i$ enhance the spectral slope, producing a local maximum (or bump) at an intermediate frequency corresponding to the scale that re-enters near such  transitions. The resulting spectrum therefore exhibits  a single or double-peaked profile depending on the number and duration of the subsequent stiff epochs.

\begin{figure}[htb]
    \centering
    {\includegraphics[width=0.485\textwidth]{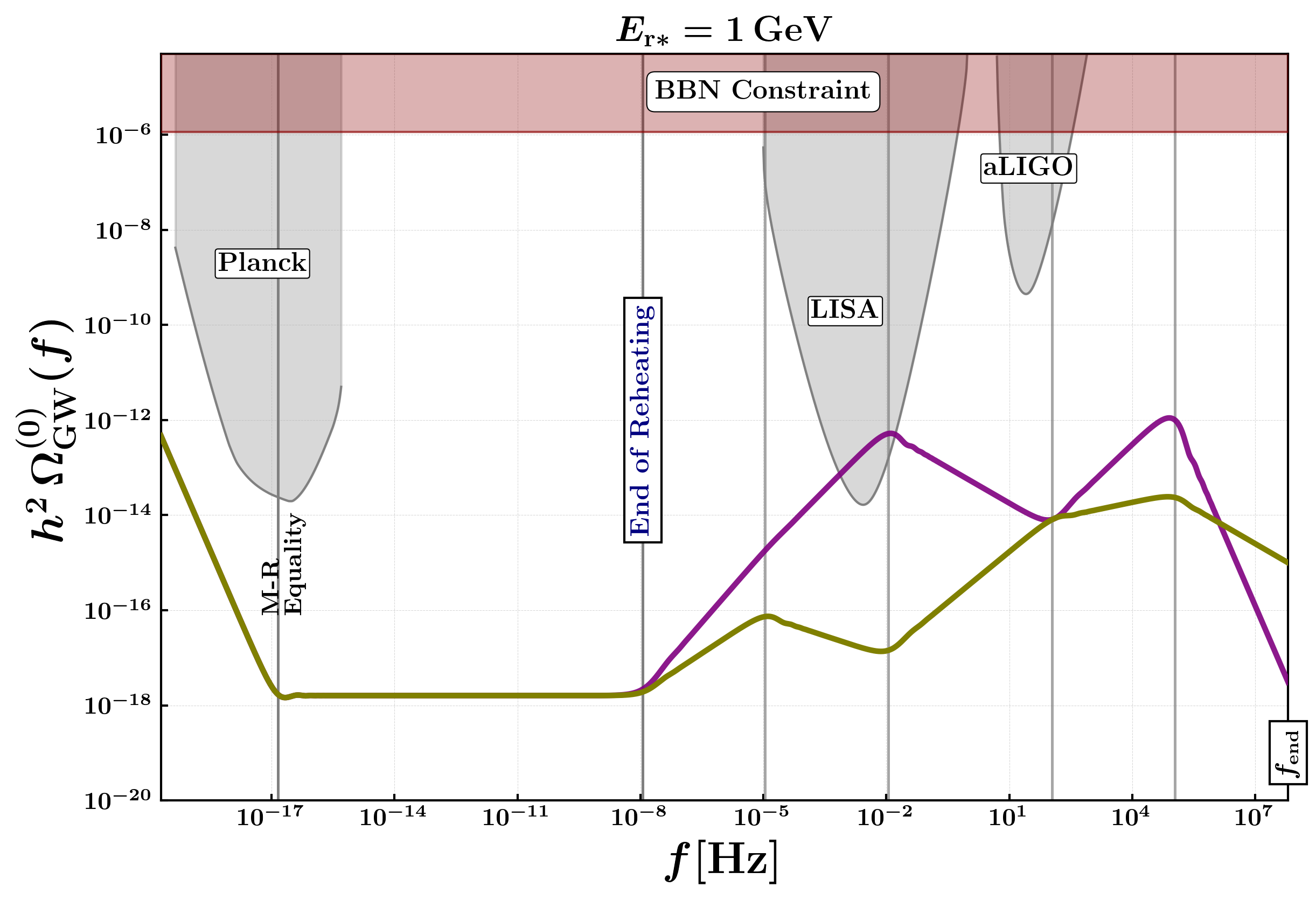}}
    {\includegraphics[width=0.485\textwidth]{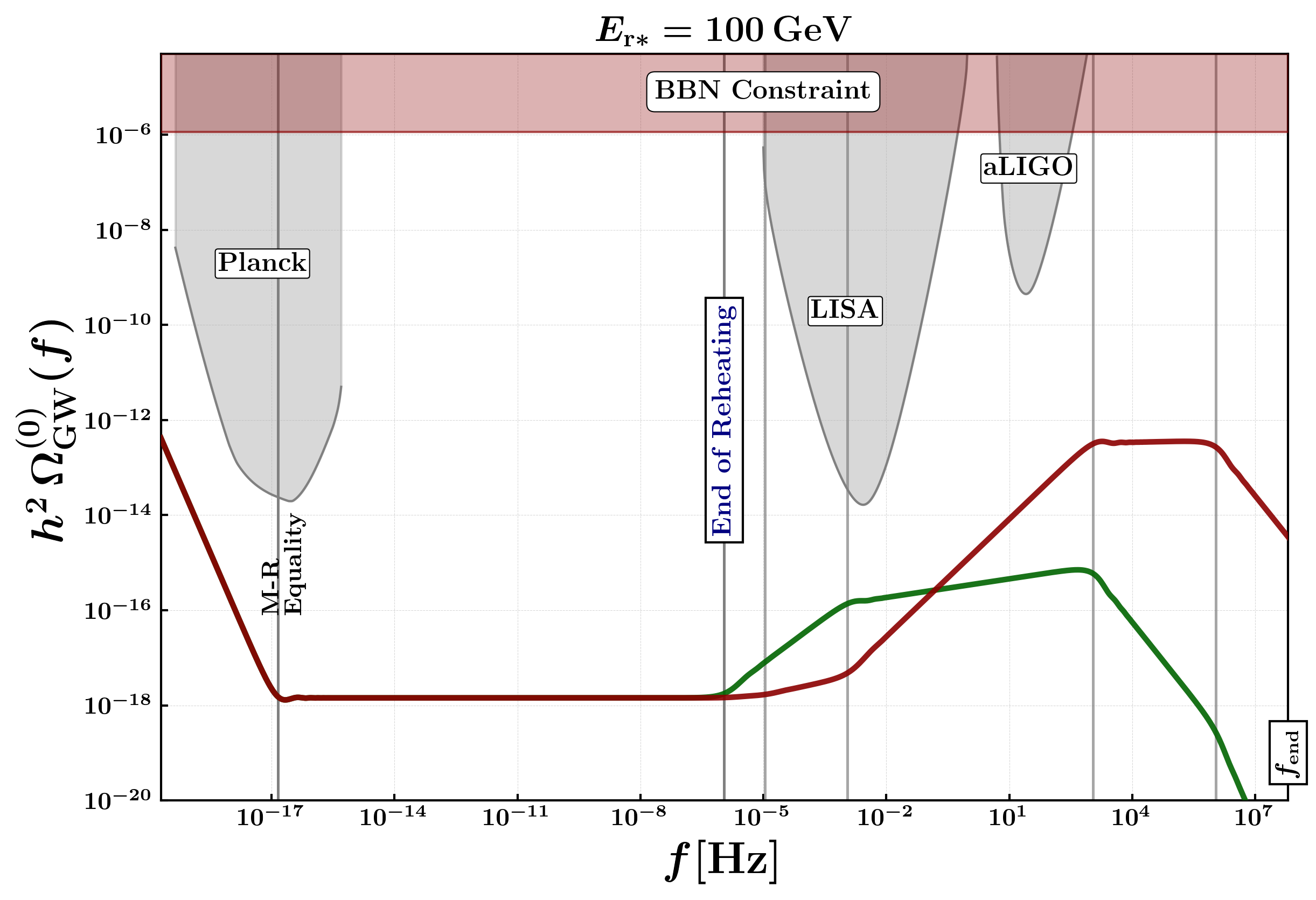}}
    {\includegraphics[width=0.485\textwidth]{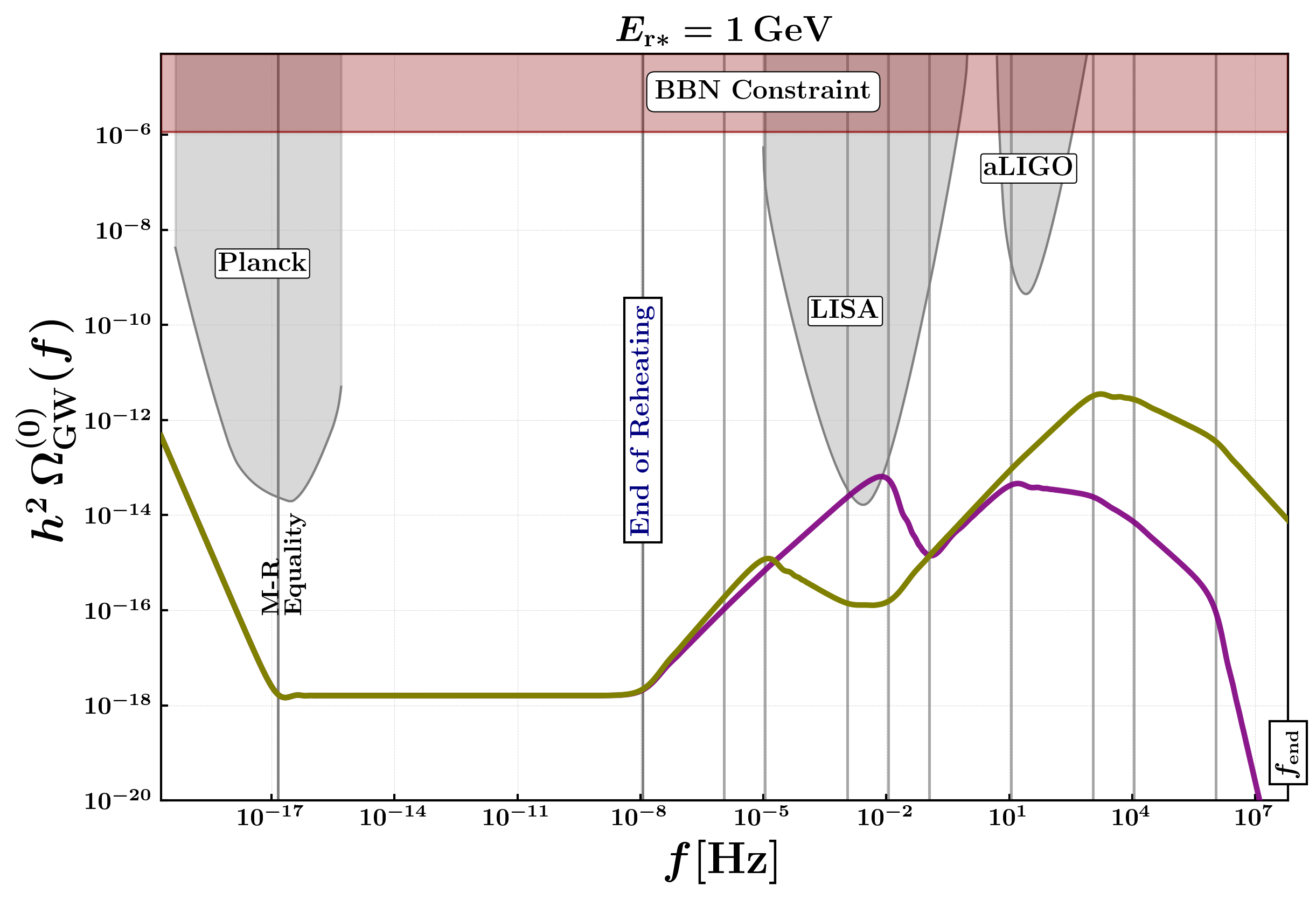}}
    {\includegraphics[width=0.485\textwidth]{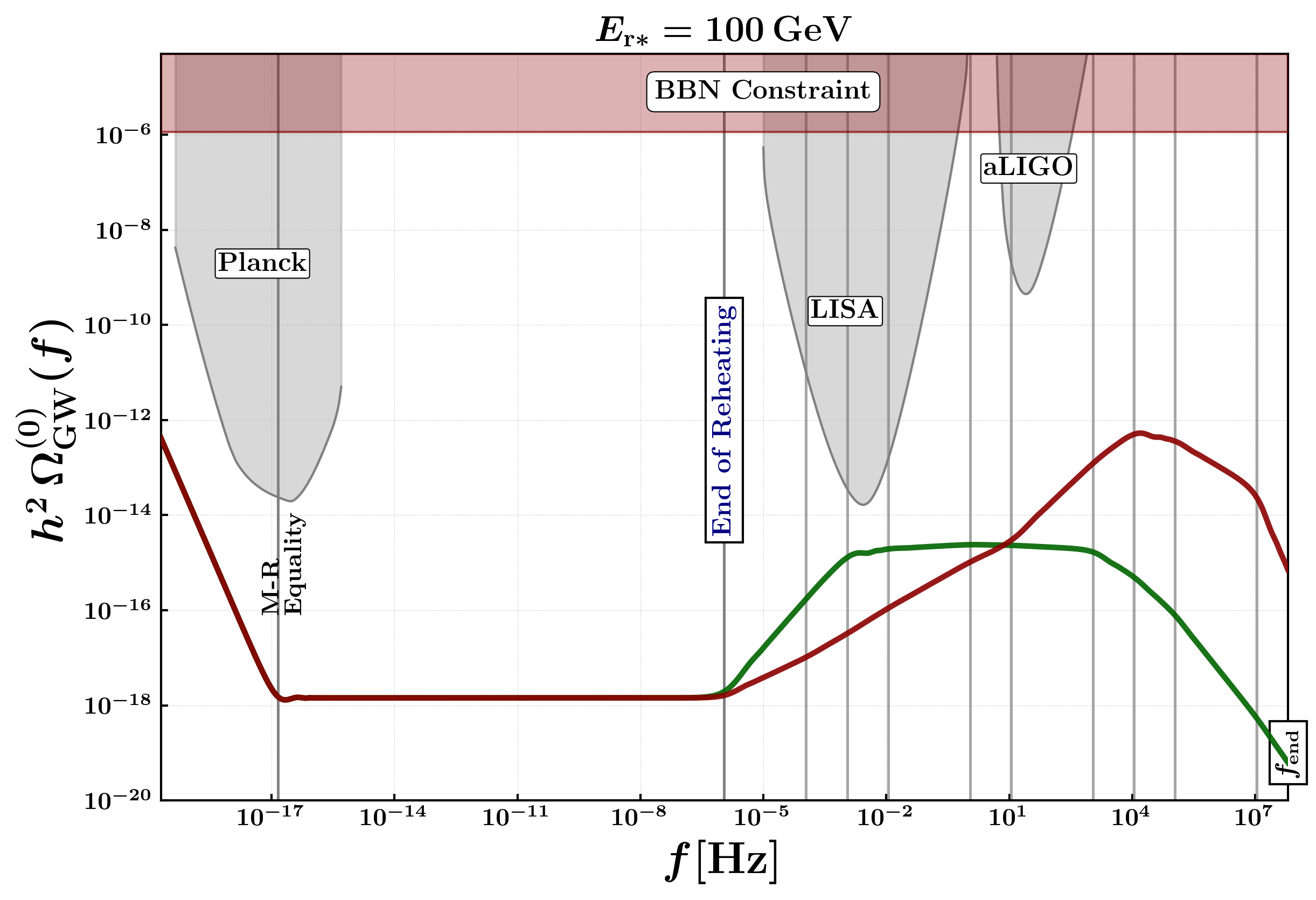}}
    \caption{
        Spectral energy density of inflationary GWs, $\Omega^{(0)}_{\rm GW}(f)$, for the case of {\bf non-monotonic spectra with peak(s)}, arising when the post-inflationary universe begins with a softer EoS $(w_1<1/3)$, but later evolves to become stiffer $(w > 1/3)$ at some epoch. The {\bf top row} contains 5 different post-inflationary epochs, while the {\bf bottom row} contains 10 epochs. The {\bf left column} corresponds to an energy scale of $E_{\rm r*}=1 \, \mathrm{GeV}$ at the beginning of hot Big Bang, while the {\bf right column} corresponds to $E_{\rm r*}=100 \, \mathrm{GeV}$. Transitions between successive epochs are marked by grey vertical lines. The BBN constraint is illustrated by the red-shaded region, and the sensitivity curves for \texttt{LISA}, \texttt{Planck}, and \texttt{aLIGO} are displayed by grey-shaded regions. [None of the curves violate the existing constraints from BBN and \texttt{aLIGO}.]
        }
    \label{fig:soft_nonmonotonic}
\end{figure}

Physically, this behaviour reflects the temporary dilution of GW energy density during the soft phase(s),  followed by a relative amplification in the stiff epoch(s). A representative class of non-monotonic GW spectra with $w_1 < 1/3$ is shown in~Fig.~\ref{fig:soft_nonmonotonic}.

\subsubsection{Non-monotonic spectra for a stiff EoS immediately after inflation}
\label{sec:Inf_GW_Zoo_nonmonotonic_stiff}
Conversely, if the universe initially enters a stiff post-inflationary epoch ($w_1 > 1/3$), including a brief kination phase, the corresponding GW spectrum exhibits an opposite sequence of features. The early blue tilt, generated by the rapid dilution of the background energy density during the stiff epoch, enhances $\Omega^{(0)}_{\rm GW}(f)$ at high frequencies, which is  associated with modes making an early horizon re-entry. 

\begin{figure}[htb]
    \centering
    {\includegraphics[width=0.485\textwidth]{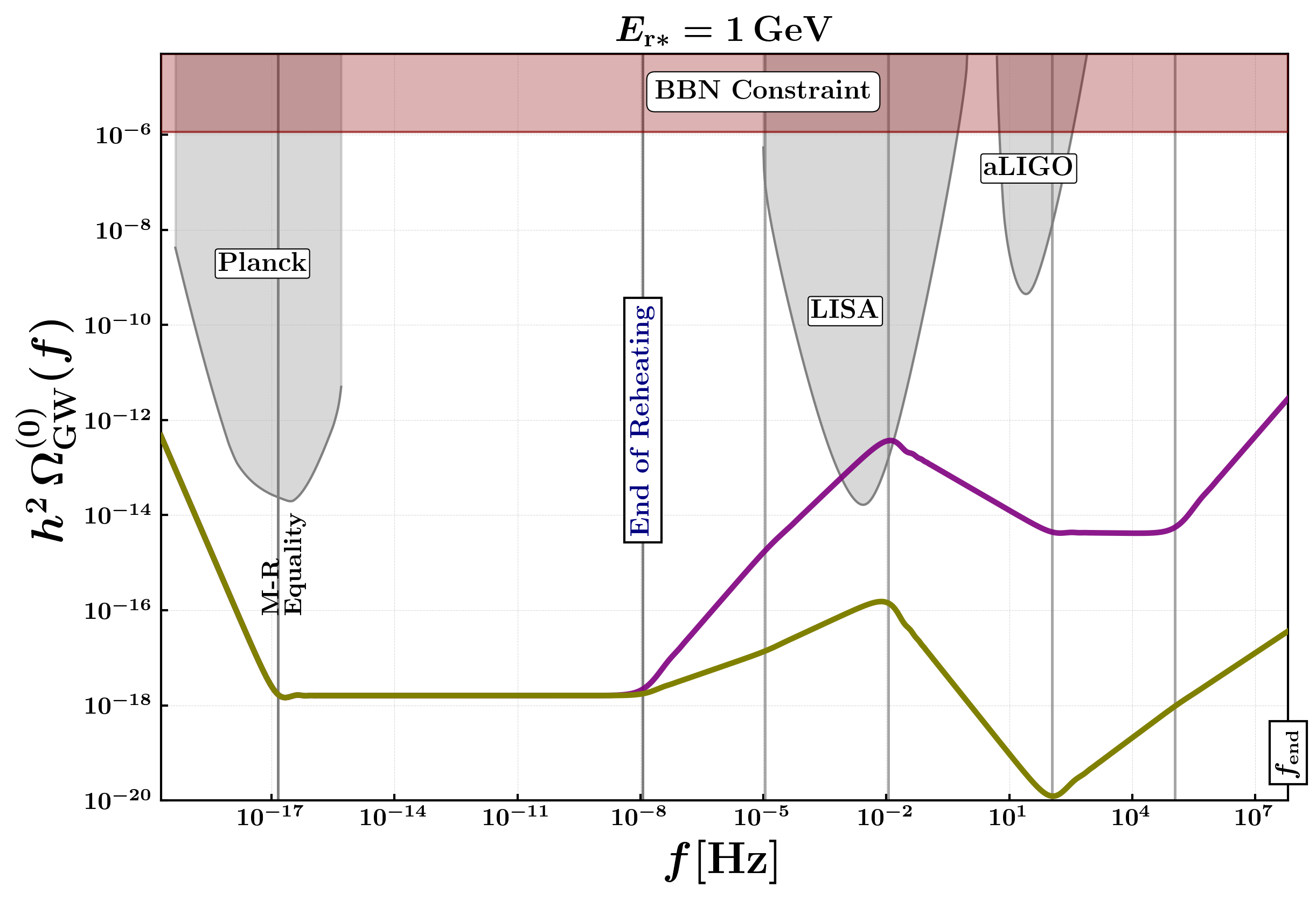}}
    {\includegraphics[width=0.485\textwidth]{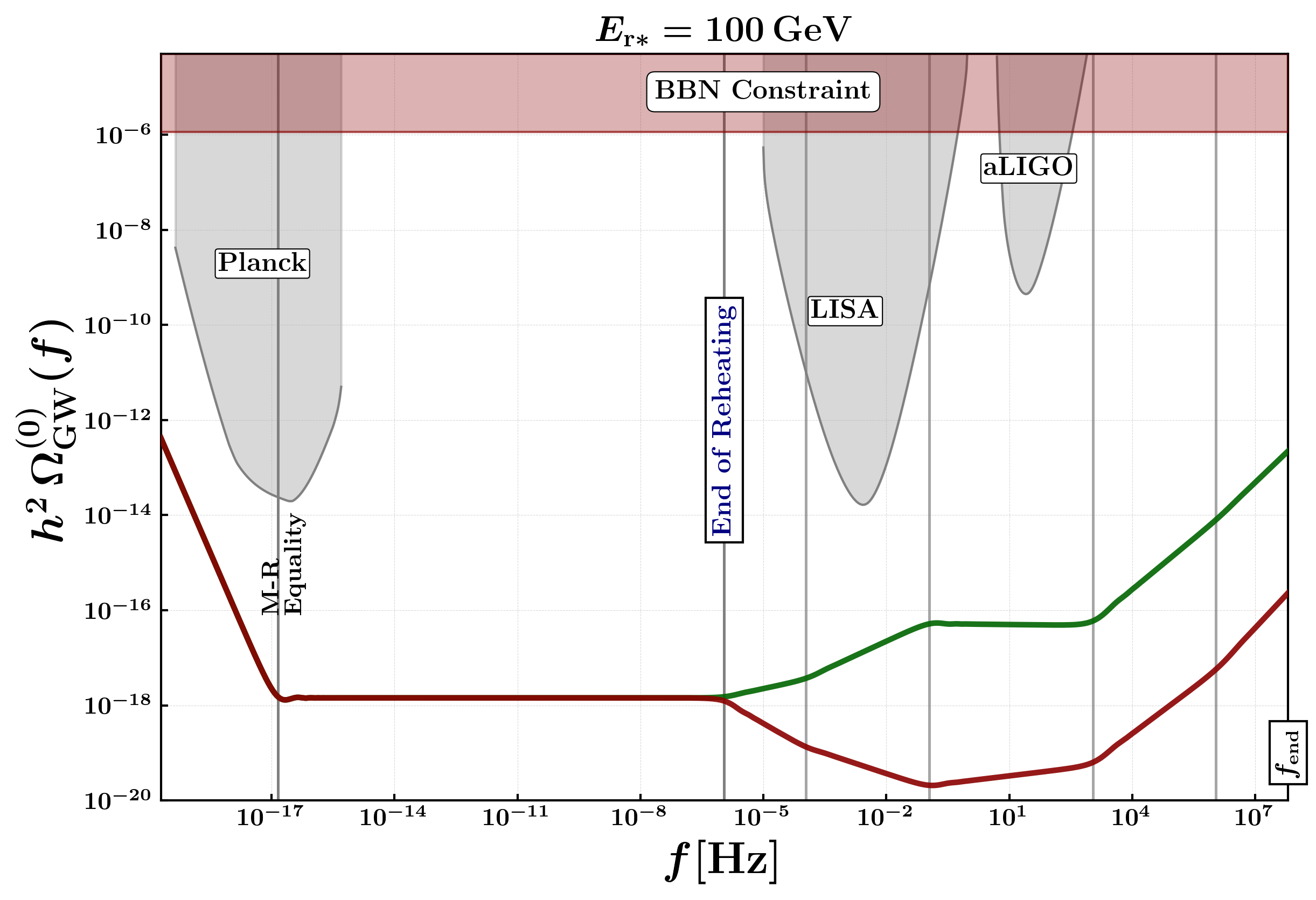}}
    {\includegraphics[width=0.485\textwidth]{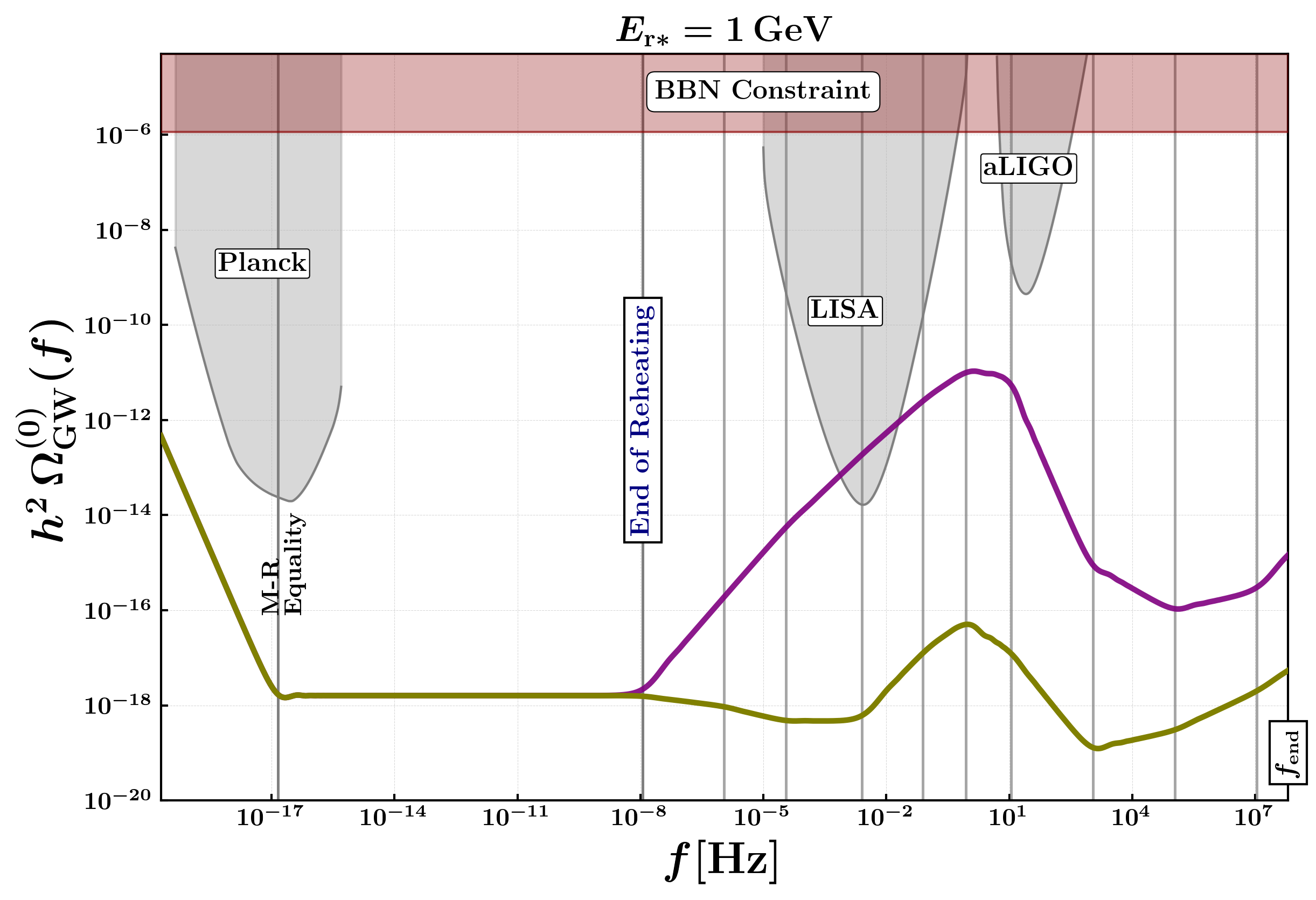}}
    {\includegraphics[width=0.485\textwidth]{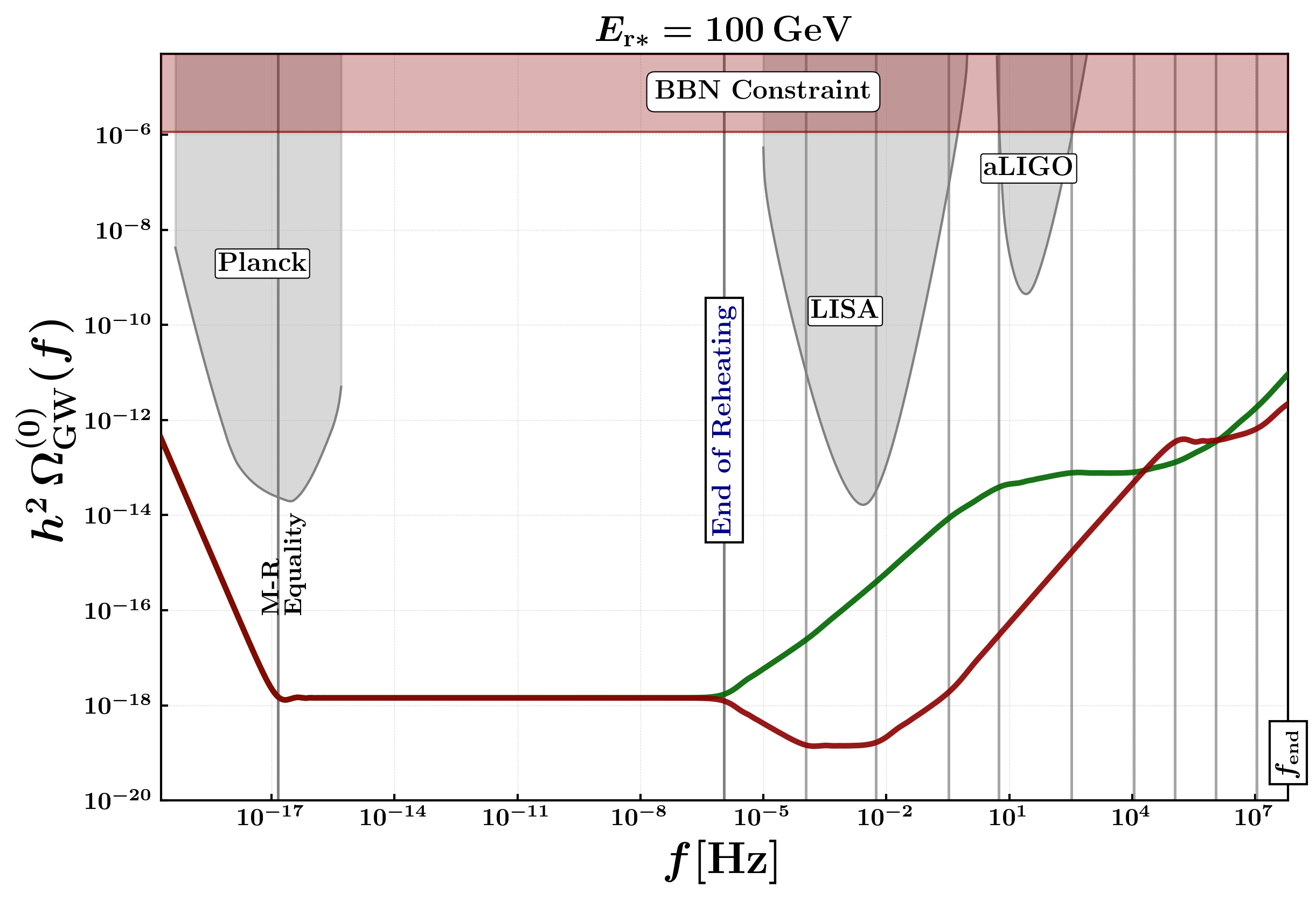}}
    \caption{
    Spectral energy density of inflationary GWs, $\Omega^{(0)}_{\rm GW}(f)$, for the case of {\bf non-monotonic spectra}, arising when the post-inflationary universe begins with a stiffer EoS $(w_1 > 1/3)$, but later evolves to become softer $(w < 1/3)$ at some epoch. The {\bf top row} contains 5 different post-inflationary epochs, while the {\bf bottom row} contains 10 epochs. The {\bf left column} corresponds to an energy scale of $E_{\rm r*}=1 \, \mathrm{GeV}$ at the beginning of hot Big Bang, while the {\bf right column} corresponds to $E_{\rm r*}=100 \, \mathrm{GeV}$. Transitions between successive epochs are marked by grey vertical lines. The BBN constraint is illustrated by the red-shaded region, and the sensitivity curves for \texttt{LISA}, \texttt{Planck}, and \texttt{aLIGO} are displayed by grey-shaded regions. [None of the curves violate the existing constraints from BBN and \texttt{aLIGO}.]
        }
    \label{fig:stiff_nonmonotonic}
\end{figure}

When the universe subsequently transitions to a softer phase ($w_{(i > 1)} < w_1$), the spectral slope decreases, producing a turnover that manifests as a suppression, some times as a trough (or dip) at intermediate frequencies. This generates a non-monotonic structure characterised by one or more peaks depending on the number of alternating stiff–soft transitions. A representative class of non-monotonic GW spectra with $w_1 > 1/3$ is shown in~Fig.~\ref{fig:stiff_nonmonotonic}.

\bigskip

 Primordial GWs with both the  aforementioned types of non-monotonic spectra, with one or more peaks,  are particularly interesting from a phenomenological perspective, since they can yield enhanced power in frequency bands accessible to space-based and/or ground-based GW detectors, while still remaining consistent with the BBN  constraints.

\begin{figure}[htb]
    \centering
    \includegraphics[width=0.85\linewidth]{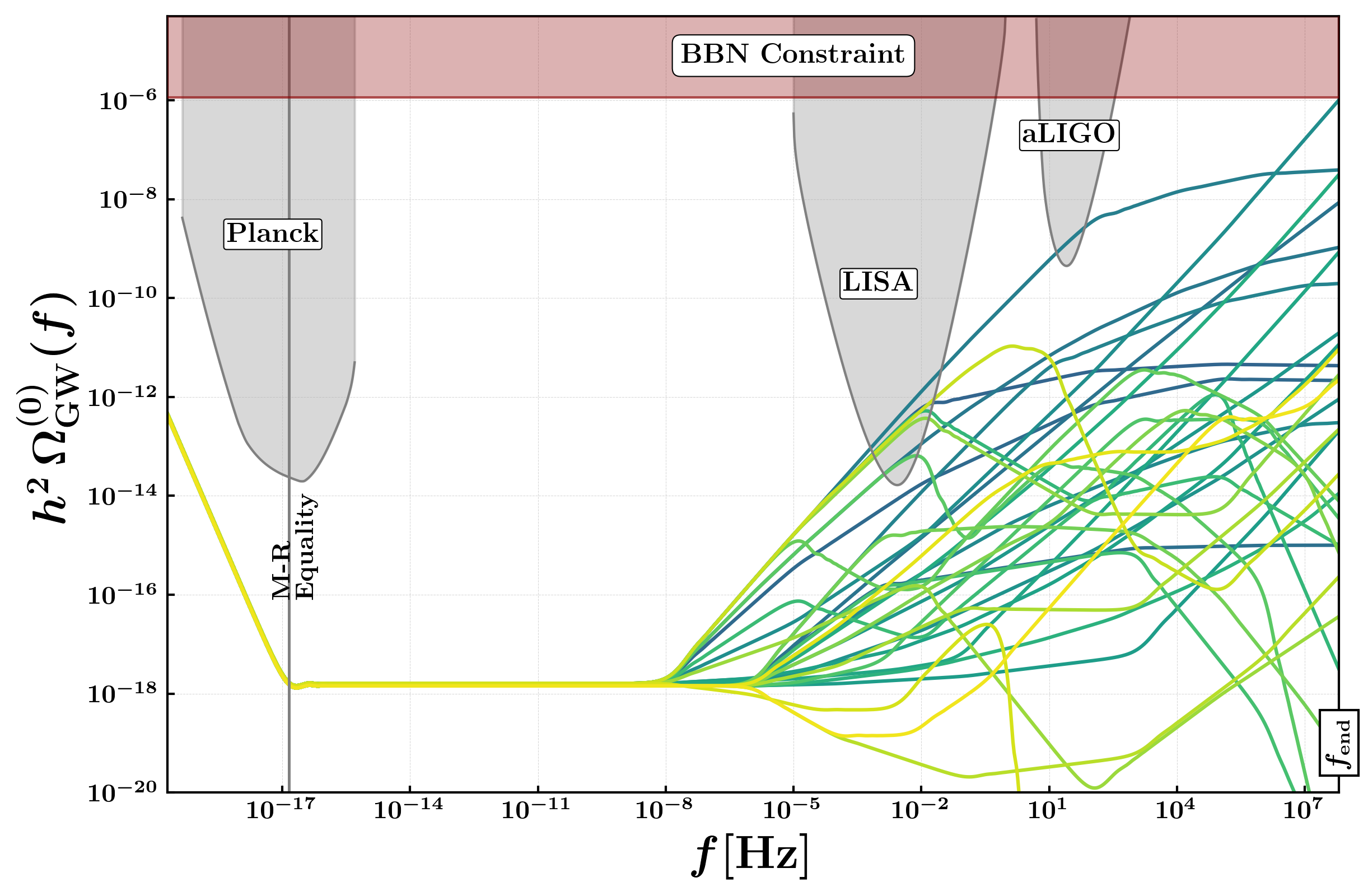}
    \caption{{\bf \Brown Spectral zoo plot} illustrating a variety of possible morphologies of the inflationary GW spectra $\Omega_{\rm GW}^{(0)}(f)$, generated by stacking together all the curves appearing in Figs.~\ref{fig:convex_monotonic}\,--\,\ref{fig:stiff_nonmonotonic} (with the colour scheme corresponding to the order in which we plotted all the curves from Fig.~\ref{fig:convex_monotonic} to Fig.~\ref{fig:stiff_nonmonotonic}, without any physical significance).}
    \label{fig:zoo}
\end{figure}

\section{\texttt{GWInSpect}:~Python-based numerical package}
\label{sec:Num_GWInSpect}

\subsection{Overview}
\label{sec:GWInSpect_Overview}

\texttt{GWInSpect}\,\href{https://github.com/athul104/GWInSpect.git}{\faGithub} is a \texttt{Python}-based numerical framework designed to compute the present-day spectral energy density of inflationary gravitational waves, $\Omega_{\rm GW}^{(0)}(f)$, for a user-specified post-inflationary expansion history consisting of multiple epochs with instantaneous transitions. Each epoch is characterized by a constant equation-of-state parameter, and the code implements the junction matching conditions described in Ref.~\cite{Soman:2024zor} to evolve the tensor modes across successive epochs.  

The package is lightweight, modular, and written for both exploratory and precision studies of the inflationary GW background. It employs the following major libraries:
\begin{itemize}
    \item \textbf{NumPy:} Array programming and vectorized numerical computing for scientific workflows~\cite{harris2020array}.
    \item \textbf{SciPy:} Used the special function Gamma~\cite{2020SciPy-NMeth}.
    \item \textbf{mpmath:} For high-precision evaluations of Bessel functions, ensuring numerical stability in oscillatory regimes~\cite{mpmath}.
\end{itemize}

\subsection{Required inputs}
\label{sec:GWInSpect_Inputs}

The following inputs are required to run \texttt{GWInSpect}. Unless otherwise stated, any input lying outside the specified range is automatically rejected with a descriptive message.

\begin{itemize}
    \item \textbf{List of EoS parameters:} 
    $\{w_1, w_2, \ldots, w_n\}$, ordered from the earliest to the latest post-inflationary epoch, satisfying $-0.28 \leq w_i < 1$.  
    \item \textbf{Transition energies:} 
    An ascending list $\{E_{n-1}, \ldots, E_1\}$ in GeV, defining the boundaries between epochs (latest $\rightarrow$ earliest).  
    The typical range is $10^{-3}\,{\rm GeV} \lesssim E_i \lesssim 10^{16}\,{\rm GeV}$.  
    The terminal hot Big Bang energy scale $E_{\rm r*}$ is not included in this list.
    \item \textbf{Commencement of hot Big Bang:} 
    Either specify the hot Big Bang temperature $T_{\rm r*} \geq 10^{-3}\,{\rm GeV}$ (consistent with BBN constraints) or  the corresponding energy scale $E_{\rm r*}$, which the code internally maps to $T_{\rm r*}$.
    \item \textbf{Inflationary scale:} 
    Specify either the tensor-to-scalar ratio $r$ (restricted to $r < 0.036$) or the corresponding energy scale of inflation, $E_{\rm inf} < 1.4\times 10^{16}\,{\rm GeV}$.
\end{itemize}

\begin{shaded}
\vspace{-0.2in}
\begin{center}
\underline{\bf \Brown Duration of each epoch.}
\end{center}
To maintain the validity of the instantaneous-transition approximation discussed in Sec.~\ref{subsec:tensor_postinf}, \texttt{GWInSpect} enforces that every post-inflationary epoch must span at least one $e$-fold. The number of $e$-folds in the $i^{\rm th}$ epoch is defined as
\begin{equation}\label{eq:efold_check}
\boxed{~\bm{\Delta N_i} \;=\; \frac{4}{3(1+\bm{w_i})}\,
\ln\!\l(\bm{\frac{E_{i-1}}{E_{i}}}\r)~} \;\geq\; 1
\quad \forall\, i\,\in\,\l\{1, 2,  \cdots ,\,n \r\}\,.
\end{equation}
If $\Delta N_i < 1$ for any epoch, the code halts execution and prompts the user to revise the duration (energy hierarchy) or the EoS parameters of the input sequence.  
For transparency, the complete array $\{\Delta N_i\}$ can also be given as one of the output for user inspection.
\end{shaded}
\subsection{Typical outputs}
\label{sec:GWInSpect_Outputs}

\texttt{GWInSpect} produces the following key outputs\,:
\begin{itemize}
    \item A log-spaced frequency array covering the range from the infrared cut-off to the UV cut-off, i.e. $f_{\rm IR}$ to $f_{\rm end}$.
    \item The corresponding present-day spectral energy density array, $\Omega_{\rm GW}^{(0)}(f)$.
    \item A list of present-day frequencies corresponding to the Hubble re-entry times of the tensor modes at each instantaneous transition (including $f_{\rm end}$ as the first entry) {\red \small (optional)}. 
    \item A list of the number of $e$-folds $\{\Delta N_i\}$ associated with each post-inflationary epoch {\red \small (optional)}.
\end{itemize}
  Together, these outputs  enable the user to reconstruct the full morphology of the inflationary GW spectra and study its dependence on the hitherto unknown expansion history of the early Universe\,\footnote{We also make available a list of useful additional functions within this package, such as the ones to check the BBN constraint using Eq.~\eqref{eq:BBN_Bound_Piecewise_Approx} and to relate different early Universe quantities at a certain epoch, amongst others.}.  We provide an instructive tutorial with an \texttt{ipynb} notebook for the reader, which can be directly accessed from the link\,:~\href{https://github.com/athul104/GWInSpect/blob/main/examples/tutorial_gwinspect.ipynb}{{\bf \blue tutorial notebook}}.

\section{Discussion and Outlook}
\label{sec:Discussions}
In this work, employing the analytical framework developed in our previous paper~\cite{Soman:2024zor}  for computing the present-day spectral energy density of first-order inflationary GWs, $\Omega^{(0)}_{\rm GW}(f)$, we  carried out a systematic investigation of their spectral shapes. A key methodological distinction of this work is its focus on the \textit{morphological diversity} of the inflationary GW spectra, including monotonic (red-tilted or blue-tilted, convex or concave) and non-monotonic (single-peaked or multi-peaked) forms, rather than on detection prospects or parameter constraints, which were the main objectives of Ref.~\cite{Soman:2024zor}.

These distinct morphologies naturally emerge from different combinations of cosmic epochs, with soft and stiff post-inflationary EoS parameters, preceding the onset of the hot Big Bang phase.  The analytical formalism developed in~\cite{Soman:2024zor}, employing Israel junction conditions and the Heaviside representation of sharp transitions, is directly applied here to study  the possible shapes of $\Omega^{(0)}_{\rm GW}(f)$ across multiple post-inflationary epochs. Apart from the BBN constraints on GWs, we enforce the condition that each post-inflationary epoch lasts at least one $e$-fold, thereby preserving the validity of the instantaneous-transition approximation underlying our analytical approach. Our analysis remains entirely model-agnostic, relying only on the sequence of EoS parameters without invoking any specific microphysical or particle-physics realization.

A complementary aim of this study is the development of a practical, user-friendly numerical package, \texttt{GWInSpect}, which automates the computation of $\Omega^{(0)}_{\rm GW}(f)$ for arbitrary sequences of the post-inflationary EoS parameters and the corresponding transition energy scales. The publicly available code provides a fast and flexible tool for generating GW spectra corresponding to a wide range of post-inflationary histories. In this way, \texttt{GWInSpect} bridges the analytical formalism of~\cite{Soman:2024zor} with numerical implementation, enabling efficient exploration of early-Universe reheating scenarios. Together, these two studies provide a simple computational framework for probing the unknown history of the early Universe through the imprints of its post-inflationary dynamics on the inflationary GW background. 

\medskip

Looking ahead, there exist several promising directions in which the present framework can be extended and refined. 

\begin{itemize}
    \item As in our previous work, the analytical framework developed across these two studies can be readily employed for model-dependent analyses, enabling one to constrain the underlying physical parameters of specific post-inflationary scenarios~\cite{Apers:2024ffe} in light of present and future GW observational prospects.
    \item By relaxing the requirement that each post-inflationary epoch must last at least one $e$-fold, through the incorporation of smooth transitions (implemented via suitable functional forms) between successive epochs, one can achieve a more realistic and fine-grained characterization of the underlying cosmological dynamics.
    \item A long-term goal is to extend the present formalism to study the impact of multiple post-inflationary transitions on the spectra of \textit{scalar-induced gravitational waves} (SIGWs)~\cite{Ananda:2006af, Baumann:2007zm, Kohri:2018awv, Domenech:2021ztg}. Since such second-order GWs depend greatly on the background expansion history, the analytical and numerical techniques developed here can offer valuable insights into their spectral structure.
\end{itemize}

\section{Acknowledgements}
We thank Mohammed Shafi and Siddharth Bhatt for valuable discussions and for carefully reading the draft and providing helpful comments. We also thank Sanket Dave for locating typos in the draft. A substantial portion of this work was carried out at IUCAA, for which SSM  gratefully acknowledges support from the IUCAA Visitor Academic Programme. Insightful discussions with Ed Copeland, Varun Sahni, Oliver Gould, Sanjit Mitra and Ranjeev Misra on gravitational waves over the years have been helpful in shaping this work. SSM was supported by the STFC Consolidated Grant [ST/T000732/1] at the University of Nottingham. AKS acknowledges funding support from SISSA as a PhD student.\\

\noindent {\bf Data Availability Statement:} This work is entirely theoretical and has no associated data.   The {\Blue\texttt{Python} package \texttt{GWInSpect}} developed to compute the spectral energy density of first-order inflationary GWs can be found in our {\Blue {\bf GitHub} repository} \href{https://github.com/athul104/GWInSpect.git}{\faGithub}, with a~\href{https://github.com/athul104/GWInSpect/blob/main/examples/tutorial_gwinspect.ipynb}{{\bf \Blue tutorial notebook}}.

\medskip

\noindent{\bf AI Assistance Statement}\,:~The authors acknowledge the use of \texttt{ChatGPT} (OpenAI GPT-5) for language editing and stylistic refinement of the manuscript.

 \medskip

For the purpose of open access, the authors have applied a CC BY public copyright license to any Author Accepted Manuscript version arising.

\newpage

\printbibliography

\end{document}